\documentclass[12pt,a4paper]{article} 
\pdfoutput=1
\usepackage{graphicx}
\usepackage{jheppub}

\usepackage{amsmath,amssymb,bm,amsfonts,yfonts}

\relax

\title{Analytic continuation of functional renormalization group equations}

\author{Stefan Floerchinger
\\
\vspace{0.1in}

Physics Department, Theory Unit, CERN, CH-1211 Gen\`eve 23, Switzerland
\vspace{0.1in}

E-mail addresses: {\tt Stefan.Floerchinger@cern.ch}
}

\abstract{Functional renormalization group equations are analytically continued from imaginary Matsubara frequencies to the real frequency axis. On the example of a scalar field with ${\cal O}(N)$ symmetry we discuss the analytic structure of the flowing action and show how it is possible to derive and solve flow equations for real-time properties such as propagator residues and particle decay widths. The formalism conserves space-time symmetries such as Lorentz or Galilei invariance and allows for improved, self-consistent approximations in terms of derivative expansions in Minkowski space.}

\begin{document}

\maketitle

\section{Introduction}
In recent years, the functional formulation of Wilsons renormalization group \cite{Wilson} has developed from exact but rather formal relations such as the flow equations of Wegner and Houghton \cite{WegnerHoughton}, Polchinski \cite{Polchinski} or Wetterich \cite{Wetterich1993} into a powerful tool for practical applications. By allowing for unconventional approximations and expansion schemes this made many interesting and partly non-perturbative investigations of quantum and statistical field theories possible. The field of applications ranges from cold atomic gases via QCD to Quantum Gravity, for reviews see \cite{FRGReviews}. 

In the modern formulation due to Wetterich one follows how a variant of the quantum effective action, the generating functional of the one-particle irreducible Feynman diagrams, changes while an infrared regulator is continuously removed. This evolution is described by an exact low equation of one-loop form. For a suitable chosen cutoff function only fluctuations of modes at a particular momentum scale $k$ contribute to the flow. This organization of the field theory with respect to the momentum scale has many advantages both for the intuitive interpretation and for practical calculations.

So far, the formalism has been mainly applied in Euclidean type space or space-time, i.\ e.\ either static, classical statistical field theories where the fields depend on spatial position only or quantum field theories in the Matsubara formalism where the time and frequencies are imaginary. In all these cases one has the important practical advantage that propagators and other correlation functions have usually at most a single isolated singularity at vanishing frequency and momentum. (The situation is more complicated for fermionic theories at non-zero density where singularities appear for example on the Fermi surface.)

In principle one can easily extend the formalism to Minkowski type space-times. Most of the definitions and derivations get modified only in a trivial way. However, in praxis a number of problems appear. Given that the invariant combinations $-p_0^2+\vec p^2$ in the relativistic case and $-p_0+\vec p^2/(2M)$ in the non-relativistic case are not positive definite, the question arises which modes actually correspond to the infrared and which to the ultraviolet. It is therefore not straight-forward to construct appropriate regulator functions, in particular if space-time symmetries such as Lorentz or Galilean boost invariance should be preserved. Also, while for Euclidean type spaces it is well understood that the flowing action approaches the microscopic or classical action for large cutoff scales, this is in general not necessarily the case in Minkowski type space-time.

Despite these problems, renormalization group methods have been applied to dynamical problems. One example are classical, stochastic, reaction-diffusion problems \cite{Canet}. This is facilitated by the fact that the microscopic inverse propagator is of the diffusion type $\partial_t - D \vec \nabla^2$ or $i\omega + D \vec p^2$ in momentum space. Thus the propagator has no pole except for $\omega=\vec p^2=0$ and one can work for example with a regulator that depends on spatial momentum, only.

A functional renormalization group formalism on the Schwinger-Keldysh closed time contour has been developed in refs.\ \cite{Gasenzer:2008zz} and \cite{Berges:2008sr}. The goal of \cite{Berges:2008sr} was the investigation of non-thermal fixed points. Many technical problems -- for example specifying a proper regulator function in Minkowski space -- could be avoided in \cite{Berges:2008sr} by integrating the flow equation and using a $1/N$ expansion to determine scaling relations between different correlation functions at a non-equilibrium fixed point. A different approach has been followed in ref.\ \cite{Gasenzer:2008zz} to access far-from-equilibrium quantum field dynamics. The regulator function was constructed with respect to time such that the renormalization group evolution and the time evolution coincide. 

Recently, functional renormalization has been used to investigate quantum gravity with Lorentz\-ian signature \cite{Manrique:2011jc}. The infrared regulator was chosen to depend on spatial derivatives only while the fluctuations in the time directions have been regulated by compactification. 
A large amount of literature exists also on applications of different renormalization group techniques to transport through a zero-dimensional quantum system such as a quantum dot which is coupled to appropriate reservoirs \cite{Zerodimensional}. 

Finally, real-time properties have been calculated from functional renormalization by analytic continuation. In refs.\ \cite{Dupuis2009} and \cite{Sinner:2009zz}, the truncated renormalization group equations for a two-dimensional gas of non-relativistic bosons have been solved in the Matsubara formalism with imaginary frequencies. The result for the propagator at the macroscopic scale $k=0$ has then been analytically continued to real frequencies using numerical Pad\'e approximant techniques. This allowed to calculate for example spectral densities and quasi-particle decay widths. A very similar approach has been followed in ref.\ \cite{SchmidtEnss} to calculate spectral functions of an impurity in a Fermi sea undergoing a transition from a polaron to a molecule.

Note that such a procedure needs generically a rather large numerical effort. The reason is that the Pad\'e approximant technique works only if enough (numerical) information about the dependence of the Propagator on the imaginary Matsubara frequency is provided. In other words, one has to calculate by some means the full frequency dependence of the (Matsubara) propagator. The numerical effort is even larger in a self-consistent scheme where the complete frequency-dependent propagator is used on the right hand side of flow equations. Nevertheless, such numerical approaches have been successfully implemented in the past, see for example ref.\ \cite{BMW} for a scheme that allows to resolve both the momentum and field dependence of the propagator. This scheme has been applied to relativistic scalar fields at non-zero temperature in ref.\ \cite{BMWScalar}.

An alternative is a non-self-consistent scheme where a simplified form of the propagator -- approximated for example by a derivative expansion -- is used on the right hand side of flow equations. In that case the interesting and non-trivial information about the spectral properties are only extracted from the flow equation but do not help to improve the results for other observables, e.\ g.\ thermodynamic potentials.

In the present paper we develop an approach to the calculation of dynamic properties from functional renormalization which deviates substantially from all others that have been followed before. We use a linear response framework where the analytic continuation from imaginary Matsubara frequencies to real frequencies is done on the level of the flow equations and not on the level of the final result at the macroscopic scale $k=0$. This has a number of advantages:

(i) Since the flow equations of objects such as the propagator are usually available in analytic form or as an integral expression one can do the analytic continuation by hand and does not have to use involved numerical techniques such as Pad\'e approximants.

(ii) The interesting and non-trivial information about the real-time properties such as quasi-particle decay widths can be used in a self-consistent way on the right hand side of flow equations. This allows to improve the performance of a truncation in general so that also properties not directly connected to the propagator  -- for example thermodynamic properties -- can be calculated with higher accuracy.

(iii) Our formalism conserves translational symmetries as well as space-time symmetries such as Lorentz or Galilei boost invariance. Due to a convenient choice of the infrared regulator function it is nevertheless possible to perform the Matsubara summations in loop expressions analytically. This leads to well behaved expressions on the right hand side of flow equations where at most an integral over spatial momenta remains to be done numerically.

(iv) In comparison with approaches based on the Schwinger-Keldysh closed time contour, the approach taken here is less involved. Also, since it is based on the well understood formalism in Euclidean space, one can profit from available knowledge. For example, it is known how the flowing action approaches the microscopic action for large cutoff scales or how useful regulator functions can be constructed.

On the other side it is clear that a formalism that is based on analytic continuation is restricted to close-to-equilibrium situations. More specific, one can only calculate what can be accessed from linear response theory. More complicated non-linear response properties as they dominate in far-from-equilibrium situations are beyond the scope of this setup.

This paper is organized as follows. In section \ref{sec:effectiveaction} we recall the definition and the analytic properties of the quantum effective action, the generating functional of the one-particle irreducible Feynman diagrams. This is followed by a discussion of the main principles of  functional renormalization in its modern formulation for Euclidean space in section \ref{sec:PrinciplesofFRGinEuclidean}. Our main results concerning the analytic continuation of flow equations are presented in section \ref{sec:analyticcontiuationofflowequations}. We first discuss the principle idea and propose a suitable class of infrared regulator functions. On the example of a scalar field with $O(N)$ symmetry we explain then the technical details and show that the formalism works well in praxis. We solve the flow equation within a truncation of the space of possible action functionals putting particular emphasis on real-time properties such as propagator residues and decay widths. Finally, we draw some conclusions in section \ref{sec:conclusions}. Appendix \ref{app:integralfunctions} collects some definitions of integral functions that are used throughout section \ref{sec:analyticcontiuationofflowequations}.

\section{The quantum effective action and its analytic structure}\label{sec:effectiveaction}
In this brief introductory section we recall the definition of the quantum effective action or one-particle irreducible effective action and discuss its properties, in particular its analytic structure. This section does not contain new results but combines existing knowledge that is usually discussed in different contexts. We use it to introduce our notation and to provide the basement for the discussion in later sections.

\subsection{The Schwinger functional}
The formalism we develop in the following can be applied to all variants of quantum field theories. In general the spectrum might contain bosonic and fermionic fields of different spin and in various representations of internal symmetries. However, in the present paper we are not interested in a particular model but in more conceptual issues. We concentrate therefore on the technically simplest case of a single scalar field. The presented formula can be generalized in a straight forward way to more complicated situations. (For fermionic Grassmann fields one needs some care in tracing the additional minus signs.) 

We are interested in the calculation of expectation values and correlation functions of the type
\begin{equation}
\int D\varphi \; {\cal O}[\varphi] \; e^{-S[\varphi]}.
\label{eq:expectvalueop}
\end{equation}
Here, the operator ${\cal O}[\varphi]$ may be a quite general functional of the field $\varphi$, for example a product of fields at different space-time points. The action $S$ is taken to be real (hermitean) as appropriate for a quantum field theory after  analytic continuation from Minkowski to Euclidean space. Usually, the action is given as an integral over a Lagrange density which depends on the fields $\varphi$ and its derivatives at the point $x$,
\begin{equation}
 S = \int_x {\cal L}(\varphi(x), \partial_\mu \varphi(x),\dots).
\end{equation}
To calculate expectation values and correlation functions as in Eq.\ \eqref{eq:expectvalueop}, one introduces the partition functional
\begin{equation}
 Z[J] = \int D \varphi \; e^{-S[\varphi] + \int_x J \varphi}.
\label{eq:partfunctfunctintegral}
\end{equation}
By taking functional derivatives, one obtains expectation values
\begin{equation}
 \phi(x) = \langle \varphi(x) \rangle = \frac{1}{Z[J]} \frac{\delta}{\delta J(x)} Z[J]
\end{equation}
as well as correlation functions
\begin{equation}
\langle \varphi(x) \varphi(y)\dots\rangle = \frac{1}{Z[J]} \frac{\delta}{\delta J(x)} \frac{\delta}{\delta J(y)} \dots Z[J].
\end{equation}
We note that these objects depend on the source field $J$. 

Connected correlation functions can be obtained more direct from the Schwinger functional
\begin{equation}
 W[J] = \ln Z[J],
 \label{eq:Schwingerfunct}
\end{equation}
for example
\begin{equation}
 \langle \varphi(x) \varphi(y) \rangle_c = \langle \varphi(x) \varphi(y) \rangle - \phi(x) \phi(y) = \frac{\delta}{\delta J(x)} \frac{\delta}{\delta J(y)} W[J].
\label{eq:connectedtwopointfunctfromW}
\end{equation}

For free theories, where $S[\varphi]$ is quadratic in the fields $\varphi$, one can usually calculate the Schwinger functional analytically. However, for non-trivial theories with interactions, this is not the case. If the interaction parameter is small, one can use a perturbative expansion. Various other approximative methods such as $1/N$ expansion or $\epsilon$-expansion exist, as well.

It is crucial that $W[J]$ encodes the information of all correlation functions. In terms of perturbation theory it contains contributions from all orders in the coupling constant as well as from all loop orders. This indicates that $W[J]$ is usually hard to determine in a closed form. On the other side, a theory is basically ``solved'' if this can be done and many observables follow from functional derivatives of $W[J]$.

For practical purposes it is often useful to work not directly with the Schwinger functional $W[J]$ but with a close relative, the quantum effective action $\Gamma[\phi]$. Its definition and properties will be reviewed in the following subsection.

\subsection{Quantum Effective action}
\label{sec:QuantumEffectiveaction}
The quantum effective action is defined as the Legendre transform of the Schwinger functional
\begin{equation}
 \Gamma[\phi] = \int_x J \phi - W[J],
\label{eq:definitionGamma}
\end{equation}
with $\phi$ the expectation value
\begin{equation}
\phi(x) = \frac{\delta}{\delta J(x)} W[J] = \langle \varphi(x) \rangle.
\end{equation}
Directly from its definition one obtains the field equation
\begin{equation}
 \frac{\delta}{\delta \phi(x)} \Gamma[\phi] = J(x).
\label{eq:fieldeqgamma}
\end{equation}
Taking another derivative, one finds
\begin{equation}
(\Gamma^{(2)}[\phi])(x,y) = \frac{\delta^2}{\delta \phi(x) \delta \phi(y)} \Gamma[\phi] = \frac{\delta J(y)}{\delta \phi(x)}  = (W^{(2)}[J])^{-1}(x,y).
\label{eq:gamma2asinverseofw2}
\end{equation}

The quantum effective action is useful for several reasons. Since it is the Legendre transform of the Schwinger functional, it contains basically the same information. This implies that correlation functions of all orders in perturbation theory can be obtained from functional derivatives of $\Gamma[\phi]$.

In perturbation theory $\Gamma[\phi]$ can also be defined as the generating functional of the one-particle irreducible Feynman diagrams. While this seems to be a quite formal statement, it has a direct practical consequence: The quantum effective action is exact at tree level. By this we mean that if the quantum effective action $\Gamma[\phi]$ is known, everything that remains to be done to calculate a physical observable such as a scattering cross section is to evaluate tree-level Feynman diagrams with the vertices and propagators taken from the quantum effective action $\Gamma[\phi]$. This implies in particular that the quantum effective action contains directly the physical parameters such as masses, charges, magnetic moments and so on as they are measured in experiments. No loop diagrams need to be calculated any more. The parameters appearing in $\Gamma[\phi]$ are already the renormalized ones. It is important to emphasize at this point that the quantum effective action contains not only vertices that are ``renormalizable'' in the sense of perturbative field theoretic renormalization theory. In contrast, it contains in general all terms that are allowed by symmetries. For example, the quantum effective action of QED will contain a term of the form
\begin{equation}
 \int_x \kappa\; \bar \psi i \sigma^{\mu\nu} \psi \; F_{\mu\nu}
\end{equation}
with a coefficient $\kappa$ that is related to the anomalous magnetic moment $g-2$ by $\kappa = \frac{-e}{4 m}(g-2)$. 

To see that $\Gamma[\varphi]$ is ``exact at tree level'' we note that the Schwinger functional can be obtained from the quantum effective action by a Legendre transform
\begin{equation}
 W[J] = \int J \phi - \Gamma[\phi]
\label{eq:WJasLegendretransformfromGamma}
\end{equation}
where $\phi$ on the right hand side is determined from the implicit equation $J=\frac{\delta}{\delta \phi}\Gamma[\phi] $. One can now see in different ways that the exact connected $n$-point functions generated by $W[J]$ can be written as tree diagrams with propagators and vertices obtained from $\Gamma[\phi]$.

One possibility is a ``proof by example''. Taking functional derivatives of $W[J]$ one finds (we use a symbolic notation for simplicity)
\begin{equation}
\begin{split}
W^{(1)} & = \frac{\delta}{\delta J_\alpha} W[J] = \phi_\alpha,\\
W^{(2)}_{\alpha\beta} & = \frac{\delta}{\delta J_\alpha} \frac{\delta}{\delta J_\beta} W[J] = \frac{\delta \phi_\beta}{\delta J_\alpha} = (\Gamma^{(2)})^{-1}_{\alpha\beta},\\
W^{(3)}_{\alpha\beta\gamma} & = - (\Gamma^{(2)})^{-1}_{\alpha\alpha^\prime} (\Gamma^{(2)})^{-1}_{\beta\beta^\prime} (\Gamma^{(2)})^{-1}_{\gamma\gamma^\prime} \Gamma^{(3)}_{\alpha^\prime\beta^\prime\gamma^\prime},\\
W^{(4)}_{\alpha\beta\gamma\delta} & = \dots
\end{split}
\label{eq:proofbyexample}
\end{equation}
By continuing this one finds indeed that $W^{(n)}$ can be written as a tree level expression involving the propagator $(\Gamma^{(2)})^{-1}$ and the vertices $\Gamma^{(n)}$. 

Another, more elegant way to see this  \cite{Coleman85, Weinberg96} 
is to consider for a moment a functional integral where $\Gamma[\varphi]$ replaces the microscopic action $S[\varphi]$
\begin{equation}
 e^{W_\Gamma[J]} = \int D \varphi \; e^{-\frac{1}{g}\Gamma[\varphi] + \frac{1}{g} \int J \varphi}.
\label{eq:functintwithgammaass}
\end{equation}
A loop expansion for $W_\Gamma[J]$ is an expansion in the parameter $g$ with
\begin{equation}
 \left(W_\Gamma[J]\right)_{L \;\text{loops}} \sim g^{L-1}.
\end{equation}
In particular, the term of order $g^{-1}$ contains only tree diagrams.  One evaluates now the functional integral in Eq.\ \eqref{eq:functintwithgammaass} in the limit $g\to 0$. It is then dominated by the first term in the saddle point approximation, the stationary phase term
\begin{equation}
 \lim_{g\to 0} \left(g W_\Gamma[J]\right) = \int J \varphi - \Gamma[\varphi],
\end{equation}
with $\varphi$ determined by $J = \delta\Gamma[\varphi]/\delta\varphi$. However, this is just the defintion of $W[J]$ in Eq.\ \eqref{eq:WJasLegendretransformfromGamma}. In other words, we found that the Schwinger functional $W[J]$ can be obtained from a ``functional integral'' as in \eqref{eq:functintwithgammaass} where the microscopic action $S$ is replaced by $\Gamma$ and only tree diagrams are taken into account.

\subsection{Perturbative loop expansion}
It is clear from the considerations in the previous section that the quantum effective action is a useful object. On the other side it is also clear that it is general not easy to calculate. To investigate the connection between $\Gamma[\phi]$ and the microscopic action $S[\varphi]$ more closely, it is useful to consider the following implicit functional integral representantion which can be derived directly from the definitions after a shift in the integration measure
\begin{equation}
 e^{-\Gamma[\phi]/\hbar} = \int D\varphi \; e^{-S[\phi+\varphi]/\hbar + \frac{1}{\hbar}\int \left(\frac{\delta}{\delta \phi} \Gamma[\phi] \right) \varphi}.
\label{eq:effectiveactionfunctintegral}
\end{equation}
We have restores here the units of $\hbar$ for a moment. One can now understand the perturbative loop expansion as an expansion of the right hand side in Eq.\ \eqref{eq:effectiveactionfunctintegral} in $\hbar$. One finds
\begin{equation}
 \Gamma[\phi] = \text{const} + S[\phi] + \hbar \frac{1}{2} \text{STr} \ln S^{(2)}[\phi] + \dots
\end{equation}
The ellipses stand for two loop and higher order loop expressions. Assuming that the terms multiplying powers of $\hbar$ on the right hand side are finite, one finds $\Gamma[\phi] = S[\phi]$ (up to an irrelevant constant) in the classical limit $\hbar\to 0$. This is the reason why $S[\phi]$ is sometimes called ``the classical action''.

\subsection{Analytic structure}
So far we have restricted the discussion to Euclidean quantum field theory (``imaginary time''). Concepts like the Schwinger functional or the quantum effective action are of course also useful for accessing real time properties. We note at this point that the definition of the effective action $\Gamma[\phi]$ can in principle be easily extended from Euclidean to Minkowski space, at least in the vacuum for vanishing temperature. The fields $\varphi$ of the functional integral depend then on the real time variable $t$ in addition to the space position $\vec x$. The correlation functions one calculates as expectation values of field monomials correspond to time-ordered correlation functions in the operator picture.

However, we take here another point of view where the configuration space for the functional integral fields $\varphi$ and their expectation values $\phi$ is always Euclidean space. (At non-zero temperature it is a generalized torus with circumference $1/T$ in the imaginary time direction.) However, the $n$-point functions which can be obtained from $\Gamma[\phi]$ as functional derivatives, can be analytically continued towards the real frequency axis. For example, from the second functional derivative of $\Gamma[\phi]$ with respect to the fields $\phi$ one can obtain the propagator $G(p_0,\vec p)$ at the imaginary Matsubara frequencies $p_0=i \omega_n$. As will be discussed below, it follows from general principles that this function can be analytically continued to the complete plane of complex frequencies except for possible poles and branch cut singularities on the real axis.

A similar analytic continuation is possible for the higher order correlation functions and in this sense the effective action in Euclidean or Matsubara space contains already all the relevant information for the dynamics in Minkowski space. It is clear, however, that within the approach propagated here one can only access close-to-equilibrium properties in the spirit of linear response theory. In contrast, a more elaborate setup based for example on the Schwinger-Keldysh formalism would be needed in a far-from-equilibrium situation. 

Let us now discuss constraints on the analytic structure of $n$-point functions as they arise from basic principles of quantum field theory in more detail. For simplicity we restrict the considerations here to the two-point function $\Gamma^{(2)}$ but we stress that at no point we assume that the fields we consider are fundamental. Therefore, by applying similar arguments to two-point correlations of composite fields one can in principle cover much of the analytic structure of a general $n$-point function $\Gamma^{(n)}$. In a homogenous situation, the two-point function can be written as
\begin{equation}
\Gamma^{(2)} (p, p^\prime) = \frac{\delta }{\delta \phi(-p)} \frac{\delta}{\delta \phi(p^\prime)} \Gamma[\phi] = (2\pi)^d \delta^{(d)} (p-p^\prime) \; G^{-1}(p)
\end{equation}
with $G(p)$ being the momentum representation of the Euclidean propagator. 

For a microscopic action $S[\varphi]$ of the standard form
\begin{equation}
S[\varphi] = \int d^dx \; {\cal L}(\varphi, \partial_\mu \varphi),
\label{eq:microscopicactionstandard}
\end{equation}
one can go from the functional integral representation to an operator picture. Translations with respect to (real) time are generated by a Hamiltonian with real and positive eigenvalues. Similarly, momentum operators with real eigenvalues generate spatial translations. The Hilbert space gets spanned by a complete set of states which can be taken to be eigenstates of the four-momentum operator $\hat P^\mu$. Using this formalism as well as requirements from space-time symmetry, causality and unitarity, one can derive the following K\"allen-Lehmann spectral representation of the propagator (for a detailed derivation see e.g. \cite{Weinberg96})
\begin{equation}
G(p) = \int_0^\infty d \mu^2 \; \rho(\mu^2) \frac{1}{p^2+\mu^2}
\label{eq:KLrepresentation}
\end{equation}
with real and non-negative spectral weight $\rho(\mu^2)\geq 0$. 

The representation \eqref{eq:KLrepresentation} is valid for an Euclidean quantum field theory where $p^2=\vec p^2+p_0^2\geq 0$ but can also be used for the analytic continuation to Minkowski signature. In Minkowski space one has to pay attention to the singularities at $p_0^2=\vec p^2 + \mu^2$ and specify an integration contour in the complex frequency plane. Often this is done by introducing additional infinitesimal terms $i\epsilon$ to move the singularities slightly away from the real $p_0$-axis along which one integrates by convention. In the representation
\begin{equation}
\int_0^\infty d \mu^2 \; \rho(\mu^2) \frac{1}{2 \sqrt{\vec p^2+\mu^2}} \left( \frac{1}{-p_0+\sqrt{\vec p^2 + \mu^2}\pm i \epsilon} - \frac{1}{-p_0-\sqrt{\vec p^2+\mu^2}\pm i \epsilon} \right)
\end{equation}
the combination $-i\epsilon$, $+i\epsilon$ in the first, respectively second term corresponds to the time-ordered propagator while the combinations $-i\epsilon$, $-i\epsilon$ represents the retarded, $+i\epsilon$, $+i\epsilon$ the advanced and $+i\epsilon$, $-i\epsilon$ the anti-time-ordered propagator. However, this $i\epsilon$-prescription is just one way of doing the book keeping and deforming slightly the contour itself is just as good.

We emphasize again that the derivation of the spectral representation \eqref{eq:KLrepresentation} works only for microscopic actions of the standard form \eqref{eq:microscopicactionstandard}. It is important that it is local and that the Lagrangian depends only on the fields $\varphi$ as well as its first derivatives $\partial_\mu \varphi$. In particular, if one adds an infrared regulator term of the form
\begin{equation}
\Delta S_k[\varphi] = \int \frac{d^d p}{(2\pi)^d} \frac{1}{2} \varphi(-p) R_k(p) \varphi(p)
\end{equation}
to the microscopic action, this corresponds in general to a non-local action, or a situation where arbitrary high derivative orders appear in the Lagrangian. It is not clear how an operator formalism can be developed in this case. One can choose the regulator function $R_k$ such that it respects space-time symmetry and causality requirements but, nevertheless, one can in general not expect that the propagator corresponding to a theory regularized in this way admits a standard spectral representation.

A few important properties can be read off from Eq.\ \eqref{eq:KLrepresentation}. First, for Euclidean signature or $p^2>0$ the propagator is real and positive. This reflects a general property of Euclidean quantum effective actions closely related to Osterwalder-Schrader reflection positivity. In contrast, close to the real frequency axis where $\text{Im}\;p_0=0$ one can write (${\cal P}$ stands for principal value)
\begin{equation}
G(p) = \int_0^\infty d \mu^2 \; \rho(\mu^2) {\cal P} \frac{1}{-p_0^2+\vec p^2+\mu^2} + i \;\pi\; \text{sign}(\text{Re}\; p_0)\; \text{sign}(\text{Im}\; p_0) \; \rho(p_0^2-\vec p^2).
\label{eq:Gpfreqaxis}
\end{equation}
This expression makes explicit that there is a branch cut on the real $p_0$ axis where the imaginary part of $G(p)$ changes its sign for those combinations of frequency and momentum where $\rho(p_0^2-\vec p^2)>0$. Away from the real frequency axis the propagator $G(p)$ is analytic and, since $\rho(\mu^2)$ cannot vanish for all $\mu^2$, also non-zero. 

In summary, the propagator $G(p)$ as a function of the complex frequency $p_0$ has poles, zero-crossings and branch-cut singularities only on the real axis. The inverse propagator $G(p)^{-1}$ has according to the above discussion all its zero-crossings, poles and branch-cut singularities on the real frequency axis, as well.

Moreover, from the representation in Eq.\ \eqref{eq:Gpfreqaxis} one can see that close to the real frequency axis one has
\begin{equation}
P(p) = G(p)^{-1} = P_1(p_0^2-\vec p^2) - i s(p_0) P_2(p_0^2-\vec p^2)
\label{eq:inversepropanalyticstructure}
\end{equation}
where we have introduced the abbreviations
\begin{equation}
s(p_0) = \text{sign}(\text{Re}\; p_0) \; \text{sign}(\text{Im}\; p_0),
\label{eq:defsp0}
\end{equation}
\begin{equation}
P_1(p_0^2-\vec p^2) = \frac{\text{Re}\; G(p)}{[\text{Re}\;G(p)]^2+[\pi \;\rho(p_0^2-\vec p^2)]^2}
\end{equation}
and $P_2(p_0^2-\vec p^2)\geq 0$ is related to $\rho(p_0^2-\vec p^2)$ by
\begin{equation}
P_2(p_0^2-\vec p^2) = \frac{\tfrac{1}{\pi} \; \rho(p_0^2-\vec p^2)}{[\text{Re}\;G(p)]^2+[\pi \;\rho(p_0^2-\vec p^2)]^2}.
\end{equation}
Close to a point $p_0^2-\vec p^2 = m^2$ where $P_1$ vanishes, one can use an expansion of the form
\begin{equation}
\begin{split}
P_1 & = Z (-p_0^2 + \vec p^2 + m^2) + \cdots\\
P_2 & = Z \gamma^2 + \cdots
\end{split}
\label{eq:230}
\end{equation}
such that for $\gamma^2 = m \Gamma$ and $\Gamma\ll m$ the propagator is of the Breit-Wigner form
\begin{equation}
G(p) = \frac{1}{Z} \frac{-p_0^2 + \vec p^2+m^2+i \; s(p_0) m \Gamma}{(-p_0^2+\vec p^2 + m^2)^2 + m^2 \Gamma^2}.
\end{equation}

The representation \eqref{eq:inversepropanalyticstructure} and expansions of the form \eqref{eq:230} will be widely used in this work; not only for the full inverse propagator $P(p)$ but also for a scale-dependent generalization thereof. Before discussing that in more detail in section \ref{sec:analyticcontiuationofflowequations}, we recall some important concepts and principles of the functional formulation of the renormalization group for Euclidean space in the next section.

\section{Principles of functional RG in Euclidean space}\label{sec:PrinciplesofFRGinEuclidean}
In this section we recall the main ideas and principles of functional renormalization for Euclidean quantum field theories. The flowing action or effective average action is introduced as a generalization of the quantum effective action. We discuss how it interpolates continuously between the microscopic or classical action and the quantum effective action while an infrared cutoff is removed. This evolution is described by an exact functional differential equation. Equally important for practical applications are approximative methods to solve it and we discuss the principle ideas behind these approximations towards the end of the section.

The present section is kept rather brief and introduces only the prerequisites for a later discussion of analytically continued flow equations in section \ref{sec:analyticcontiuationofflowequations}. For a more detailed discussion of functional renormalization in Euclidean space and applications to many concrete problems we refer to many good review articles in the literature \cite{FRGReviews}.

\subsection{Exact flow equation}
We start by modifying the partition functional and Schwinger functional introduced in \eqref{eq:partfunctfunctintegral} and \eqref{eq:Schwingerfunct}, respectively, according to
\begin{equation}
Z_k[J] = e^{W_k[J]} = \int D \varphi \; e^{-S[\varphi] - \Delta S_k[\varphi] + \int J \varphi}
\end{equation}
where we have added an infrared cutoff term
\begin{equation}
\Delta S_k[\varphi] = \frac{1}{2} \int_p \varphi(-p) R_k(p) \varphi(p)
\end{equation}
to the microscopic action $S[\varphi]$. The function $R_k(p)$ is real and positive for Euclidean momentum argument $p$. It can be of different form in different situations and we discuss possible choices below. As a general feature, one requires that $R_k(p)$ serves as an infrared cutoff as the scale $k$ which is usually implemented by the property
\begin{equation}
\lim_{p\to 0} R_k(p) \approx k^2.
\end{equation}
In addition, $R_k(p)$ should suppress fluctuations on all momentum scales for very large values of the cutoff scale $k$. This requires $R_k(p)$ to diverge for $k\to \infty$. Usually this divergence is taken is taken as strong as $k^2$. Finally, the cutoff function should vanish for $k\to 0$ for all $p$,
\begin{equation}
\lim_{k\to 0} R_k(p) = 0.
\end{equation}

It is now convenient to define a modified form of the effective action. The flowing action or effective average action $\Gamma_k[\phi]$ is defined by subtracting from the Legendre transform
\begin{equation}
\tilde \Gamma_k[\phi] = \int J \phi - W_k[J]
\label{eq:}
\end{equation}
with $\phi = \tfrac{\delta}{\delta J} W_k[J]$ the cutoff term
\begin{equation}
\Gamma_k[\phi] = \tilde \Gamma_k[\phi] - \Delta S_k[\phi].
\end{equation}

Using the above definitions, one can obtain an exact functional renormalization group equation for the flowing action $\Gamma_k[\phi]$, the Wetterich equation \cite{Wetterich1993}
\begin{equation}
 \partial_k \Gamma_k[\phi] = \frac{1}{2} \text{Tr} (\Gamma_k^{(2)}[\phi]+R_k)^{-1} \partial_k R_k.
\label{eq:Wetterich}
\end{equation}
We note that the right hand side has the form of a one-loop expression. Nevertheless, it is an exact equation taking all orders of perturbation theory as well as non-perturbative effects into account. 

Other properties of $\Gamma_k[\phi]$ follow from an implicit functional integral representation
\begin{equation}
 e^{-\Gamma_k[\phi]} = \int D \varphi \; e^{-S[\phi+\varphi]- \Delta S_k[\varphi]+\int (\frac{\delta}{\delta \phi}\Gamma_k[\phi])\varphi}.
\label{eq:functintrepofGammak}
\end{equation}
Similar to \eqref{eq:effectiveactionfunctintegral} this equation follows directly from the definitions together with a shift in the integration measure. 

For $k\to\infty$ and therefore $R_k\to\infty$ one finds that the cutoff term $\Delta S_k$ dominates the functional integral over the fluctuating field $\varphi$ in \eqref{eq:functintrepofGammak}. It leads to a strong suppression for large values of $\varphi(-p)\varphi(p) = \varphi^*(p)\varphi(p)$ and a saddle point expansion becomes a good approximation. One has therefore for very large $k=\Lambda$
\begin{equation}
 \Gamma_\Lambda[\phi] = \text{const} + S[\phi] + \frac{1}{2} \text{Tr} \ln (S^{(2)}[\phi]+R_\Lambda).
\label{eq:limitlargekGammak}
\end{equation}
For many applications the constant ($\phi$-independent) part is not important. 

In addition, the contribution from the one-loop term (the second term on the right hand side of \eqref{eq:limitlargekGammak}) is suppressed by negative powers of  $R_\Lambda$ for derivatives of \eqref{eq:limitlargekGammak} with respect to the field $\phi$. This implies that the flowing action effectively approaches the microscopic action for large values of the cutoff parameter $k$,
\begin{equation}
 \lim_{k\to\infty}\Gamma_k[\phi] = S[\phi].
\end{equation}

To summarize, the flowing action $\Gamma_k[\phi]$ has been defined as a modified Legendre transform of the Schwinger functional $W_k[J]$ in presence of the infrared cutoff function $R_k(p)$. For very large values of $k$ it approaches the microscopic action $S[\phi]$ (up to an irrelevant constant as well as possibly one-loop terms) and for $k=0$ it equals the quantum effective action $\Gamma[\phi]$. In other words, the flowing action flows from the microscopic action $S[\phi]$ to the quantum effective action $\Gamma[\phi]$.

The exact functional differential equation for the flowing action \eqref{eq:Wetterich} together with equation \eqref{eq:limitlargekGammak} can be seen as a differential formulation of the functional integral formalism. As an alternative formulation of quantum field theory this has an advantage for some formal questions but the most important application is for concrete practical and often approximate calculations. We will briefly discuss such an application for Euclidean field theory below. The emphasis is on the main principles of the approximation scheme rather than on the numerical results. It is useful to have these principles for Euclidean space in mind when the analytic continuation to Minkowski space is discussed thereafter. 

\subsection{Derivative expansion}
The example we discuss here is a classical scalar field theory with ${\cal O}(N)$ symmetry in $d$ spatial dimensions (usually one has $d=3$ but also $d=2$ or $d=1$ and even a single point with $d=0$ are of experimental relevance). The microscopic action $S$ is of the form
\begin{equation}
S = \int d^d x \left\{ \frac{1}{2} \vec \nabla \phi_n \vec \nabla \phi_n + \frac{1}{2} m^2 \phi_n \phi_n + \frac{1}{8} \lambda \left(\phi_n  \phi_n\right)^2 \right\}.
\label{eq:3}
\end{equation}
This model can be used for example to describe the critical properties of many second order phase transitions. 

The cutoff term for the investigation of the model in Eq.\ \eqref{eq:3} has the form (with $\phi_n^*(\vec p) = \phi_n(-\vec p)$ and $\int_p = \int \frac{d^d p}{(2\pi)^d}$)
\begin{equation}
\Delta S_k[\phi] = \int_p \frac{1}{2} \phi_n^*(\vec p) R_k(p) \phi_n(\vec p).
\label{eq:}
\end{equation}
The function $R_k(p)$ depends only on the magnitude $p=\sqrt{\vec p^2}$ so that the cutoff term is invariant under rotational symmetry. Note that it respects also other symmetries as the internal $O(N)$ symmetry and translation.
Useful cutoff functions  are for example the exponential cutoff
\begin{equation}
R_k(p) = \frac{Z_k p^2}{e^{p^2/k^2}-1},
\label{eq:expcutoff}
\end{equation}
or the one proposed by Litim \cite{Litim2001}
\begin{equation}
R_k(p) = Z_k (k^2-p^2) \theta(k^2-p^2).
\label{eq:Litimcutoff}
\end{equation}
Here $Z_k$ is the wavefunction renormalization constant and one has $Z_\Lambda=1$. 
As a feature of both functions in Eqs.\ \eqref{eq:expcutoff} and \eqref{eq:Litimcutoff} we note $R_k(p=0) = Z k^2$ and $R_k(p)\to 0$ for $p^2/k^2\to\infty$. One can see these functions as a $p$-dependent mass term for  the modes with $p^2\lesssim k^2$.

That $R_k(p)$ goes to zero for large values of $\vec p^2$ has a number of practical advantages. For a fixed value of $k$, the fact that $R_k(p) \approx 0$ for $p^2\gg k^2$ means that fluctuations of modes with large momentum are integrated in the flowing action $\Gamma_k$, already. If $k$ is lowered further, there will be no change of the correlation functions due to fluctuations of these modes. In praxis, this means that the momentum integrals that appear on the right hand side of the flow equation of an object such as the effective potential have no (or only small) contribution from momenta with $p^2\gg k^2$. There is an effective UV cutoff due to the term $\partial_t R_k(p)$ on the right hand side of the flow equation.

This picture can also be confirmed from the functional integral expression for the flowing action
\begin{equation}
e^{-\Gamma_k[\phi]} = \int D \varphi\; e^{-S[\phi+\varphi]- \int_p \frac{1}{2} \varphi_n^*(p) R_k(p) \varphi_n(p) + \int_p \frac{\delta\Gamma_k[\phi]}{\delta\phi_n(p)} \varphi_n(p)}.
\label{eq:functint1}
\end{equation}
For a given value of $k$, the modes of the fluctuation field $\varphi_n(p)$ with $p^2\gg k^2$ are not affected by the cutoff term. In contrast, for the modes with $p^2\ll k^2$ the cutoff term leads to an additional Gaussian suppression similar to a mass gap of size $k^2$.

With a cutoff function as in Eq.\ \eqref{eq:expcutoff} or \eqref{eq:Litimcutoff}, the contribution of long-ranging modes with small values of $p^2$ are included in a steady way in the functional integral in Eq.\ \eqref{eq:functint1}. These are precisely the modes that become important close to second order phase transitions and dominate the universal critical dynamics. A cutoff function that decays fast enough for large values of $p^2/k^2$ allows to separate the treatment of these modes from the higher momentum modes which are not relevant for the critical phenomena.

The above described separation of cutoff scales realized by a cutoff function $R_k(p)$ that decays sufficiently fast for large $p$ makes an expansion in terms of spatial momenta a useful approximation scheme for many purposes. In position space this corresponds to a derivative expansion. One writes
\begin{equation}
\Gamma_k[\phi] = \int d^d x \left\{ U_k(\rho) + \frac{1}{2} Z_k(\rho) \vec \nabla \phi_a \vec \nabla \phi_a + \frac{1}{4} Y_k(\rho) \vec \nabla \rho \vec \nabla \rho + {\cal O}(\partial^4)\right\},
\label{eq:derivativeexpansionrelativisticscalars}
\end{equation}
with $\rho = \tfrac{1}{2} \phi_a \phi_a$. The lowest level only includes the scalar potential and a standard kinetic term. The first correction includes the $\rho$-dependent wave function renormalizations $Z_k(\rho)$ and $Y_k(\rho)$. The next level involves then invariants with four derivatives and so on. 

On first sight there is no reason why a derivative expansion as in Eq.\ \eqref{eq:derivativeexpansionrelativisticscalars} should be a good approximation. In principle both small and large momenta contribute on the right hand side of the flow equation and even observables at small momenta could be affected by the large momenta regime where \eqref{eq:derivativeexpansionrelativisticscalars} is expected to converge badly. To understand why it nevertheless works rather well in praxis consider the following generalization of the scheme\begin{equation}
\begin{split}
\Gamma_k[\phi] =  \int d^d x {\bigg \{} & U_k(\rho) + \frac{1}{2} \vec \nabla \phi_a \, Z_k(\rho, -\vec \nabla^2) \, \vec \nabla \phi_a \\
& + \frac{1}{4} \vec \nabla \rho \, Y_k(\rho, -\vec \nabla^2)\, \vec \nabla \rho + \dots {\bigg \} }.
\end{split}
\label{eq:derivativeexpansionrelativisticscalarsimproved}
\end{equation}
The functions $Z_k$ and $Y_k$ depend now also on the momentum $p=\sqrt{\vec p^2}$. The scheme in \eqref{eq:derivativeexpansionrelativisticscalarsimproved} has the advantage that it allows to resolve the full momentum dependence of the propagator. However, in praxis one usually neglects terms higher than quadratic in the momenta. Nevertheless, derivative expansion often leads to quite good results. The reason is the following. On the right hand side of the flow equation the cutoff insertion $R_k(p)$ in the propagator $(\Gamma^{(2)}(p)+R_k(p))^{-1}$ suppresses the contribution of the modes with small momenta. On the other side, the cutoff derivative $\partial_k R_k(p)$ suppresses the contribution of very large momenta provided that $R_k(p)$ falls of sufficiently fast for large $p$. Effectively mainly modes with momenta of the order $k^2$ contribute.
 
On the other side, if one evaluates flow equations for objects as $Z_k(\rho,\vec p^2)$, $Y_k(\rho,\vec p^2)$ etc., one main effect of the external momentum is to provide an infrared cutoff scale of order $p^2$. Such an infrared cutoff scale is already provided by $R_k$ itself and one might therefore also work with the $k$-dependent but $p^2$-independent couplings
\begin{equation}
Z_k(\rho, p^2=0), \quad Y_k(\rho, p^2=0).
\end{equation}
This brings us back to \eqref{eq:derivativeexpansionrelativisticscalars}. In other words, a properly choosen cutoff function improves a truncation with $p$-independent coefficients $Z_k$ and $Y_k$ such that some of the momentum dependence is effectively taken into account. We emphasize that it is important that the cutoff $R_k(p)$ falls off sufficiently fast for large $p$. If this is not the case, the derivative expansion might lead to erroneous results since the kinetic coefficients as appropriate for small momenta and frequencies are then also used for large momenta and frequencies. Only when the scale derivative $\partial_k R_k(p)$ provides for a sufficient ultraviolet cutoff does the derivative expansion work properly.

\section{Analytic continuation of flow equations}\label{sec:analyticcontiuationofflowequations}
In this section we discuss flow equations in Minkowski space. The original formulation of the formalism as outlined in section \ref{sec:PrinciplesofFRGinEuclidean} has been in Euclidean space or for imaginary times. There are now different conceivable strategies to obtain results about real-time propagators which will be discussed below.

Our general point of view is that the flowing action $\Gamma_k[\phi]$ is {\itshape a priori} defined in the Matsubara formalism, i.\ e.\ within a functional integral setup where the configuration space for the fields $\varphi$ and their expectation value $\phi$ has a cyclic imaginary time direction with circumference $1/T$. The correlation functions that follow from $\Gamma_k[\phi]$ as functional derivatives can then be analytically continued from the discrete, imaginary Matsubara frequencies to the complex frequency plane and in particular towards the real frequency axis. 

For some properties of the flowing action the analytic continuation has no effect. This is the case in particular for the effective potential which is obtained from the flowing action by evaluating it for fields that are constant in space and time so that analytic continuation is trivial. On the other side, for frequency- and momentum-dependent objects such as the propagator or other correlation functions, it does make a difference. They can either be evaluated in Euclidean space with imaginary frequency argument or -- after analytic continuation to real frequencies -- in Minkowski space.

In most situations one is finally interested in macroscopic quantities for $k=0$. One way to get for example the macroscopic propagator in Minkowski space  would be to solve its flow equation with an arbitrary imaginary value $q_0=i \omega$ of the frequency argument (at finite temperature the frequency is restricted to the discrete values $q_0=i 2 \pi T n$ in the Matsubara formalism) and to do the analytic continuation only at the end, i.e. for $k=0$. The advantage of this procedure is that the flow equation formalism is only used in Euclidean space where it is rather transparent and well understood. On the other side, the analytic continuation can be very difficult in praxis since the propagator in Euclidean space is usually available at best numerically. Methods based for example on Pad\'e approximants need information from many Matsubara frequencies and the numerical effort gets rather large. Nevertheless, this approach has been successfully followed in the past and allowed to calculate for example the spectral function of non-relativistic bosons in two spatial dimensions \cite{Dupuis2009, Sinner:2009zz}.

In this paper we propose another way to solve the problem, however. Instead of doing the analytic continuation only after solving the flow equation or at $k=0$, we will analytically continue the flow equations themselves. Since in contrast to their solution the flow equations are usually available in analytic form, it is not necessary to use involved numerical techniques. 
As an example we investigate in particular the the flow equation for the propagator. For particular choices of the infrared cutoff function $R_k$ we show how it is possible to evaluate this object for real frequency argument and to solve the flow equations directly in Minkowski space. No analytic continuation of the final result is then needed any more to access the real time properties. Moreover, it will become apparent that such a procedure also has some advantages in designing approximations. A derivative expansion in Minkowski space can be done as a Taylor expansion around the frequency corresponding to a pole or branch-cut singularity of the propagator. Loop expressions that appear on the right hand side of flow equations but also on-shell properties of the effective action itself are strongly dominated by these singular structures. One can therefore expect that the convergence properties of such an expansion in Minkowski space are better than those of a derivative expansion around vanishing values of the frequency as it can be done in Euclidean space. 

This section is organized as follows. In subsection \ref{ssec:Regulator} we discuss a class of regulator functions $R_k$ for which the analytic continuation of the flow equation to real frequencies can be done, at least for truncations based on derivative expansion. We apply this method subsequently to a model for scalar fields with ${\cal O}(N)$ symmetry. The truncation for this particular system is introduced in subsection \ref{ssec:Truncation} while subsections \ref{ssec:Effectivepotential} and \ref{ssec:Propagator} discuss the flow equations for the effective potential and propagator, respectively. Subsection \ref{sec:Numericalresults} presents some numerical results for the solution of the flow equations within the truncation.

\subsection{Regulator}\label{ssec:Regulator}
In this subsection we discuss the particular choice of the infrared regulator function $R_k$ that allows for analytic continuation of the regularized propagator $(P_k+R_k)^{-1}$ from the imaginary frequency axis or Euclidean space to real frequencies. The discussion will be based on truncations where the inverse propagator is close to the real frequency axis of the form
\begin{equation}
P_k= Z \left( z(- p_0^2 + \vec p^2 ) + m^2 - i s(p_0) \gamma^2 \right),
\label{eq:A1}
\end{equation}
where $Z$, $z$, $m^2$ and $\gamma^2$ are $k$-dependent, real and positive quantities. 

Let us now choose a regulator function $R_k(p_0,\vec p)$. It is {\itshape a priori} not clear what requirements must be fulfilled when choosing a cutoff function in Minkowski space. For example, it is not clear what the conditions for $R_k(p_0,\vec p)$ are to be a good infrared regulator nor is it obvious what is required for it to have good regulating properties in the ultraviolet. Actually, it is not even obvious which modes correspond to the IR or UV, respectively, given that the combination $-p_0^2+\vec p^2$ is not positive definite.

On the other side, all these issues are well understood for Euclidean signature, see the discussion in section \ref{sec:PrinciplesofFRGinEuclidean}. The approach we follow here is therefore to choose a regulator function with all the desired properties in Euclidean space or for positive argument $-p_0^2+\vec p^2>0$ and to use analytic continuation to extend the calculations to Minkowski space.

For many choices of $R_k$ such a program would be rather difficult to perform since a function that is smooth and regular on the imaginary frequency axis may nevertheless have poles and discontinuities in other regions of the complex frequency plane. One can even argue that this is unavoidable if one wants $R_k(p_0,\vec p)$ to be a function that decays with large imaginary values of $p_0$. This in turn is of course necessary for $R_k$ to serve as an effective UV regulator on the right hand side of the flow equation. 

To make the analytical continuation possible in praxis we choose a class of rather simple IR regulators with algebraic decay in the UV:
\begin{equation}
R_k(p_0,\vec p) = Z k^2 \frac{1}{1+c_1\left(\frac{-p_0^2+\vec p^2}{k^2}\right) + c_2 \left(\frac{-p_0^2+\vec p^2}{k^2}\right)^2+\dots}.
\label{eq:A2}
\end{equation}
The coefficient $Z_k$ can be chosen conveniently. Usually it is taken to agree with the wave function renormalization constant but more generally it can be some real and positive function of the scale parameter $k$ with not too strong dependence on $k$.

The function \eqref{eq:A2} has all the desired properties for Euclidean arguments $-p_0^2+\vec p^2\geq 0$ if the coefficients $c_j$ are real and positive. It is clear that the UV regulating properties become better if some $c_j$ with large $j$ are non-zero. On the other side, practical calculations are simple if only a few $c_j$ with small $j$ are non-zero. Arguably the simplest, non-trivial choice is $c_1=c>0$, $c_2=c_3=\dots=0$ and we will discuss it in more detail below \footnote{A suitable cutoff function for non-relativistic field theories is obtained by replacing $-p_0^2+\vec p^2$ with $-p_0+\vec p^2/(2M)$ in equation \eqref{eq:A2}.}. One should keep in mind, however, that the decay of $R_k$ for large momenta and Matsubara frequencies is  rather mild for this choice. For the convergence properties of the widely used derivative expansion it may be desirable to have a quicker decay, either with a higher power or exponential in the momentum argument.

In the following we discuss the analytic structure of the regularized propagator $(P_k+R_k)^{-1}$ as a function of a complex frequency argument $p_0$. In principle, this depends on the complete functional form of $P_k$ and it is not sufficient to know the form of $P_k$ for example close to the real frequency axis. On the other side, at least when $k$ is sufficiently small one expects the relevant properties of the propagator to be dominated by the on-shell singularities, i.\ e.\ the poles and branch cuts on the real frequency axis. Following this rationale, we use the expression \eqref{eq:A1} not only close to the real frequency axis but everywhere in the complex $p_0$ plane, keeping in mind that the coefficients $Z$, $m^2$ and $\gamma^2$ should be determined at the singularity, i.\ e.\ close to the point on the real frequency axis where the real part of $P_k$ vanishes.

With the above choice of the regulator function it is straightforward to show that the regularized propagator can be written as (we use $\tilde m^2=m^2/k^2$ and $\tilde\gamma^2=\gamma^2/k^2$)
\begin{equation}
(P_k+R_k)^{-1} = \frac{1}{Z} \left( \frac{\beta_1}{p^2 + \alpha_1 k^2} + \frac{\beta_2}{p^2+\alpha_2 k^2} \right)
\label{eq:A3}
\end{equation}
with
\begin{equation}
\begin{split}
\alpha_{1/2} & = \frac{1}{2}\left(\frac{1}{c}+ \frac{\tilde m^2}{z}-i\,s(p_0)\, \frac{\tilde \gamma^2}{z} \right) \pm \left[A+ i\, s(p_0)\, B\right] ,\\
\beta_{1/2} & = \frac{1}{2 z} \pm \left[ C + i\; s(p_0)\, D \right]
\end{split}
\label{eq:alphabeta}
\end{equation}
and 
\begin{equation}
\begin{split}
A &= \frac{1}{2i} \left[ \sqrt{\frac{1}{c z} - \frac{1}{4}\left( \frac{1}{c} - \frac{\tilde m^2}{z}-i \frac{\tilde \gamma^2}{z} \right)^2} -\sqrt{\frac{1}{c z} - \frac{1}{4}\left( \frac{1}{c} - \frac{\tilde m^2}{z}+i \frac{\tilde \gamma^2}{z} \right)^2}  \right],\\
B &= \frac{1}{2} \left[ \sqrt{\frac{1}{c z} - \frac{1}{4}\left( \frac{1}{c} - \frac{\tilde m^2}{z}-i \frac{\tilde \gamma^2}{z} \right)^2}+\sqrt{\frac{1}{c z} - \frac{1}{4}\left( \frac{1}{c} - \frac{\tilde m^2}{z}+i \frac{\tilde \gamma^2}{z} \right)^2} \right]\\
C &= \frac{-A \left( \tfrac{1}{c}-\frac{\tilde m^2}{z} \right) - B \,\frac{\tilde \gamma^2}{z}}{4\, z\, (A^2+B^2)},\\
D &= \frac{B \left( \tfrac{1}{c}-\frac{\tilde m^2}{z} \right) - A\, \frac{\tilde \gamma^2}{z}}{4\, z\,(A^2+B^2)}.
\end{split}
\end{equation}
We choose the branch cut of the square root function to be along the negative real axis. The objects $A$, $B$, $C$ and $D$ are then always real. Moreover, using the above definitions it is straightforward to establish that for all values of $c>0$, $\tilde m^2\geq 0$ and $\tilde \gamma^2\geq 0$ one has
\begin{equation}
\begin{split}
& \frac{1}{2}\left(\frac{1}{c}+\frac{\tilde m^2}{z}\right) \pm A \geq 0,\\
& B \geq 0,\\
& \frac{\tilde \gamma^2}{z} - B \leq 0.
\end{split}
\end{equation}

The decomposition in \eqref{eq:A3} is very useful for practical calculations since it resembles closely the form of a free propagator. A similar representation is actually possible for the whole class of regulators in \eqref{eq:A2}.

We note at this point that in contrast to the propagator $P_k^{-1}$ in Eq.\ \eqref{eq:A1}, the regularized propagator $(P_k+R_k)^{-1}$ in \eqref{eq:A3} has singularities away from the real frequency axis. Indeed, for $\frac{\tilde \gamma^2}{z}-B<0$ and $s(p_0)=1$ there are poles at $p_0=\pm \sqrt{\vec p^2+\alpha_1 k^2}$. A spectral representation as in Eq.\ \eqref{eq:KLrepresentation} is therefore not possible for $(P_k+R_k)^{-1}$. Although we do not give a proof here we believe that this is a generic feature for cutoff functions that serve as an effective UV regulator.

One of the most important features of the regulator function proposed here is that one can calculate in praxis with the decomposition \eqref{eq:A3}. Besides the pole singularities, $(P_k+R_k)^{-1}$ has also a branch cut originating from the second term in \eqref{eq:A3}. However, in an approximation where one assumes that all integrals along this branch cut are dominated by the nearby poles on the different Riemann sheets one can always perform the frequency integrals (or Matsubara summations for $T>0$) analytically which is an enormous advantage for practical calculations.

\subsection{Truncation for $O(N)$ model of scalar fields}\label{ssec:Truncation}
In the remainder of this section we perform the analytical continuation for the flow equations of a concrete model, the $O(N)$ model for a scalar field in the phase with spontaneous symmetry breaking. The excitation spectrum consists of a massive radial mode as well as $N-1$ massless Goldstone modes. Due to the non-linear coupling, the radial mode can actually decay into two Goldstone excitations which gives rise to a non-vanishing decay width.

Since the breaking of the $O(N)$ symmetry is spontaneous, the effective action and its $k$-dependent generalization, the flowing action $\Gamma_k[\phi]$ are invariant with respect to global $O(N)$ symmetry transformations. It must also be invariant under translational and rotational symmetries as well as under Lorentz transformations. Apart from these constraints it may be a rather general functional of the field $\phi$. However, to make a (numerical) solution of the flow equations possible it is in any case necessary to truncate the space of possible functionals $\Gamma_k[\phi]$. 

In this paper we use a truncation based on derivative expansion in Minkowski space. More specific, we take $\Gamma_k$ to be of the form
\begin{equation}
\begin{split}
\Gamma_k  = \int_{t,\vec x} {\Bigg \{}  & \sum_{M=1}^N \frac{1}{2} \bar \phi_M \, \bar P_\phi(i\partial_t,-i\vec \nabla) \, \bar \phi_M\\
& + \frac{1}{4} \bar \rho\, \bar P_\rho(i\partial_t, -i\vec \nabla) \, \bar \rho + \bar U_k(\bar \rho) {\Bigg \} }
\end{split}
\label{eq:truncationbar}
\end{equation}
with $\bar \rho = \frac{1}{2} \sum_{M=1}^N \bar \phi_M^2$. We fix some arbitrariness in this decomposition by demanding the momentum dependent parts to vanish for $p_0=\vec p =0$, i.e.
\begin{equation}
\bar P_\phi(0,0) = \bar P_\rho(0,0) = 0.
\end{equation}

We now make the crucial assumption that the momentum dependent propagator parts $\bar P_\phi(p_0,\vec p)$ and $\bar P_\rho(p_0,\vec p)$, considered as a function of a complex frequency $p_0$, are of the same analytic structure as the full inverse propagator in \eqref{eq:inversepropanalyticstructure}. This is certainly the case at the macroscopic scale $k=0$ since $\Gamma_{k=0}[\phi] = \Gamma[\phi]$. For non-zero $k$ one expects in principle deviations due to the frequency dependence of the cutoff function $R_k(p_0,\vec p)$. One can see it as an element of our truncation that these deviations are neglected, however.

In this spirit and taking constraints from Lorentz invariance into account we write the functions $\bar P_\phi$ and $\bar P_\rho$ as
\begin{equation}
\bar P_{\phi/\rho}(p_0,\vec p) = \bar Z_{\phi/\rho}(-p_0^2+\vec p^2) \left[ -p_0^2+\vec p^2 \right] - i s(p_0) \bar \gamma_{\phi/ \rho}^2 (-p_0^2+\vec p^2). 
\end{equation}
We use again the function $s(p_0)$ defined in Eq.\ \eqref{eq:defsp0}. The functions $\bar \gamma_{\phi/\rho}^2(-p_0^2+\vec p^2)$ are nonzero only for negative argument. 

To derive the flow equation for $\bar U_k(\bar \rho)$, we expand \eqref{eq:truncationbar} around a background configuration according to
\begin{equation}
\bar \phi_1(t,\vec x) = \bar \phi + \delta \bar \phi_1(t,\vec x), \quad \bar \phi_M(t,\vec x) = \delta \bar \phi_M(t,\vec x) \quad (M=2,\ldots, N)
\label{eq:expansionaroundconstantfields}
\end{equation}
and keep only terms that are quadratic in the ``fluctuating parts'' $\delta \bar \phi_M$. This yields ($\bar \rho = \frac{1}{2}\bar \phi^2$)
\begin{equation}
\begin{split}
\Gamma_{k,2} = \int_{\vec p,p_0} & \frac{1}{2}\delta \bar \phi_1(-p_0,-\vec p) {\big [} \bar P_\phi(p_0,\vec p) + \bar \rho \bar P_\rho(p_0,\vec p)  + \bar U_l^\prime(\bar \rho) + 2 \bar \rho \bar U_k^{\prime\prime}(\bar \rho) {\big ]} \delta \phi_1(p_0,\vec p) \\
& + \sum_{M=2}^N \frac{1}{2} \delta \bar \phi_M(-p_0,-\vec p) {\big [} \bar P_\phi(p_0,\vec p) + \bar U_k^\prime(\bar\rho) {\big ]} \delta \bar \phi_M(p_0,\vec p)
\end{split}
\label{eq:Gamma_2trunc}
\end{equation}
In principle one could now derive a closed set of flow equations for the set of functions $\bar Z_\phi(-p_0^2+\vec p^2)$, $\bar Z_\rho(-p_0^2+\vec p^2)$, $\bar \gamma_\phi^2(-p_0^2+\vec p^2)$, $\bar \gamma_\rho^2(-p_0^2+\vec p^2)$ and $\bar U_k(\bar \rho)$ by using appropriate projections of the flow equation for $\Gamma_k$. At the macroscopic scale $k=0$, the functions $\bar Z_\phi(-p_0^2+\vec p^2)$, $\bar Z_\rho(-p_0^2+\vec p^2)$, $\bar \gamma_\phi^2(-p_0^2+\vec p^2)$ and $\bar \gamma_\rho^2(-p_0^2+\vec p^2)$ would contain all the interesting information of the propagator, spectral density etc. However, this procedure would result in a set of coupled, partial integro-differential equations which is rather hard to solve numerically.

Instead we therefore devise here a simpler approximation based on the observation that the physical properties of a propagator are to a very large extend dominated by singularities, in particular poles or branch cuts corresponding to zero-crossings of the inverse propagator or at least its real part. Around these singularities it is sensible to use a derivative expansion or Taylor expansion of the inverse propagator with respect to its frequency and momentum arguments.

We observe at this point that the frequency $p_0$ where the propagators become singular depend on the value of the background field $\bar\rho$. The most important properties will be dominated by the minimum $\bar \rho_0$ of the effective potential where $\bar U^\prime(\bar\rho_0)=0$, however. We therefore choose for the inverse propagator of the Goldstone modes the expansion point $p_0=\vec p=0$. The imaginary part $\bar\gamma_\phi^2$ vanishes at this point and we write
\begin{equation}
\bar P_\phi(p_0,\vec p) \approx \bar Z_\phi(0) (-p_0^2+\vec p^2),
\end{equation}
the higher orders being neglected.

The situation is different for the radial field. Since it is massive, it is sensible to choose the expansion point as $p_0=m_1$ and write there
\begin{equation}
\begin{split}
& \bar P_\phi(p_0,\vec p) + \bar \rho_0 \bar P_\rho(p_0,\vec p)\\
& \approx\left[ \bar Z_\phi(-m_1^2) + \bar \rho_0 \bar Z_\rho(-m_1^2) \right] (-p_0^2+\vec p^2) - i s(p_0) \left[ \bar \gamma_\phi^2(-m_1^2) + \bar \rho_0 \bar \gamma_\rho^2(-m_1^2) \right],
\end{split}
\label{eq:413}
\end{equation}
with higher order terms being neglected.

We define now the wave function renormalization as $\bar Z_\phi=\bar Z_\phi(0)$, the renormalized fields as $\phi_M=\sqrt{\bar Z_\phi} \bar \phi_M$, $\rho=\bar Z_\phi \bar \rho$, the effective potential $U_k(\rho)= \bar U_k(\bar \rho)$ and the anomalous dimension as $\eta_\phi = - \frac{1}{\bar Z_\phi} \partial_t \bar Z_\phi$. In addition we introduce the abbreviations $Z_1$ and $\gamma_1^2$ by
\begin{equation}
\begin{split}
Z_1 &= \frac{1}{\bar Z_\phi} \left( \bar Z_\phi(-m_1^2) + \bar \rho_0 \bar Z_\rho(-m_1^2) \right)\\
\gamma_1^2 &= \frac{1}{\bar Z_\phi} \left( \bar \gamma_\phi^2(-m_1^2) + \bar \rho_0 \bar \gamma_\rho^2(-m_1^2) \right).
\end{split}
\label{eq:defZ1gamma1}
\end{equation}
An expression analogous to \eqref{eq:Gamma_2trunc} but in terms of renormalized quantities can now be written
\begin{equation}
\begin{split}
\Gamma_{k,2} = \int_{\vec p,p_0} & \frac{1}{2}\delta \phi_1(-p_0,-\vec p) \left[ Z_1 \left[-p_0^2+\vec p^2 \right] - i s(p_0) \gamma_1^2 + U_k^\prime(\rho) + 2 \rho U_k^{\prime\prime}(\rho) \right] \delta \phi_1(p_0,\vec p) \\
& + \sum_{M=2}^N \frac{1}{2} \delta \phi_M(-p_0,-\vec p) \left[ -p_0^2+\vec p^2 + U_k^\prime(\rho) \right] \delta \phi_M(p_0,\vec p)
\end{split}
\label{eq:Gamma_2truncren}
\end{equation}
One can directly read off the (inverse) propagators for the excitation spectrum from this expression by evaluating it for $\rho=\rho_0$. Since $U_k^\prime(\rho_0)=0$, the $N-1$ Goldstone bosons are massless and have a vanishing decay width as expected. In contrast, the radial mode $\phi_1$ has a mass $m_1=\sqrt{2\rho_0 U_k^{\prime\prime}(\rho_0)/Z_1}$ and a decay width $\Gamma_1 = \gamma_1/\sqrt{2\rho_0 U_k^{\prime\prime}(\rho_0)}$.

\subsection{Flow of the effective potential}\label{ssec:Effectivepotential}
Using \eqref{eq:Gamma_2truncren} it is straight-forward to derive a flow equation of $U_k(\rho)$. To that end one simply evaluates equation \eqref{eq:Wetterich} for constant argument $\phi=\sqrt{2 \rho}$. The result is 
\begin{equation}
\begin{split}
\partial_t U_k(\rho) {\big |}_\rho & = \frac{\partial}{\partial \bar \rho} U_k(\rho) \partial_t \bar \rho{\big |}_{\rho} + \partial_t U_k(\rho) {\big |}_{\bar \rho}\\
& = \eta_\phi\, \rho\; U^\prime_k(\rho) +  \partial_t U_k(\rho) {\big |}_{\bar \rho}
\end{split}
\label{eq:B6}
\end{equation}
with
\begin{equation}
\begin{split}
\partial_t U_k(\rho){\big |}_{\bar \rho} = \frac{1}{2} \int_{p_0=i \omega_n, \vec p} & {\bigg \{} \frac{1}{Z_1 (\vec p^2 - p_0^2) - i\, s(p_0) \gamma_1^2 + U^\prime + 2 \rho U^{\prime\prime} + \tfrac{1}{\bar Z_\phi} R_k}\\
& +  \frac{(N-1)}{\vec p^2-p_0^2 + U^\prime + \tfrac{1}{\bar Z_\phi}R_k} {\bigg \}} \frac{1}{\bar Z_\phi} \partial_t R_k.
\end{split}
\label{eq:B7}
\end{equation}
In principle, the frequencies in \eqref{eq:B7} are summed over the discrete, imaginary values $p_0=i \omega_n = i 2\pi T n$ with $\int _{p_0=i \omega_n}= T \sum_n$ as appropriate for the Matsubara formalism. One should not take the expression in \eqref{eq:B7} literally at these imaginary frequency, however. In fact the expansion in \eqref{eq:413} was made around a point on the real frequency axis and the function $s(p_0)$ is only defined there. In praxis one performs the Matsubara summation in expressions like \eqref{eq:B7} usually using methods based on contour integrals in the complex frequency plane. At the end one has to evaluate residues and integrals along branch cuts on the real frequency axis (or close to it for $k>0$) where the expansion in \eqref{eq:413} and the expression in \eqref{eq:B7} can be used. 

In terms of the integral functions defined in appendix \ref{app:integralfunctions} one can write \eqref{eq:B7} as
\begin{equation}
\begin{split}
\partial_t U(\rho) {\big |}_{\bar \rho} = & \frac{k^d}{2} {\Bigg [} I_0\left( \frac{U^\prime+2\rho U^{\prime\prime}}{k^2}, \frac{\gamma_1^2}{k^2},Z_1, \frac{T}{k}, c, d \right)\\
& +(N-1)\; I_0\left( \frac{U^\prime}{k^2}, 0,1, \frac{T}{k},c,d \right) {\Bigg ]}.
\end{split}
\label{eq:flowofUintermsofintfunc}
\end{equation}

Up to terms that are proportional to $Z_\rho$ or $\gamma_\rho^2$ (and therefore subleading), the derivatives of \eqref{eq:flowofUintermsofintfunc} with respect to $\rho$ yield
\begin{equation}
\begin{split}
\partial_t U^\prime(\rho){\big |}_{\bar \rho} = \frac{k^{d-2}}{2} {\Bigg [} & \left( 3U^{\prime\prime}+2\rho U^{(3)}\right)  I_1\left( \frac{U^\prime + 2 \rho U^{\prime\prime}}{k^2}, \frac{\gamma_1^2}{k^2}, Z_1, \frac{T}{k}, c, d \right)\\
&+(N-1) U^{\prime\prime} I_1\left( \frac{U^\prime}{k^2}, 0,1, \frac{T}{k} , c, d \right) {\Bigg ]}
\end{split}
\end{equation}
and
\begin{equation}
\begin{split}
\partial_t U^{\prime\prime}(\rho){\big |}_{\bar \rho} =  \frac{k^{d-4}}{2} {\Bigg [} & - \left( 3U^{\prime\prime}+2\rho U^{(3)} \right)^2  I_2\left( \frac{U^\prime + 2 \rho U^{\prime\prime}}{k^2}, \frac{\gamma_1^2}{k^2}, Z_1, \frac{T}{k}, c, d \right)\\
& -(N-1) \left(U^{\prime\prime}\right)^2 I_2\left( \frac{U^\prime}{k^2}, 0, 1,\frac{T}{k} , c, d \right)\\
& + k^2 \left( 5 U^{(3)}+2\rho U^{(4)}\right) I_1\left( \frac{U^\prime + 2 \rho U^{\prime\prime}}{k^2}, \frac{\gamma_1^2}{k^2},Z_1 ,\frac{T}{k}, c, d \right)\\
& + k^2 (N-1) \; U^{(3)} I_1\left( \frac{U^\prime}{k^2}, 0,1, \frac{T}{k} , c, d \right) {\Bigg ]}.
\end{split}
\label{eq:B10}
\end{equation}

In principle one could attempt a solution of the flow equation \eqref{eq:B7} allowing for a completely general form of $U_k(\rho)$. Numerically one would have to solve a two-dimensional partial differential equation for that purpose. Since we are interested in conceptual and qualitative rather than precise quantitative results in the present paper, we restrict ourselves to a Taylor expansion of $U_k(\rho)$ in the form
\begin{equation}
U_k(\rho) = U_k(\rho_0) + m_k^2 (\rho-\rho_0) + \frac{1}{2} \lambda_k (\rho-\rho_0)^2,
\label{eq:trucU}
\end{equation}
the higher orders being neglected for simplicity. In the phase with spontaneous breaking of the $O(N)$ symmetry, the ($k$-dependent) minimum $\rho_{0,k}$ is at positive values, $\rho_{0,k}>0$ and the linear coefficient $m_k^2$ vanishes by definition.

In summary, our truncation consists of the $k$-dependent renormalized coefficients $Z_1$, $\gamma_1^2$, $\lambda_k$ and $\rho_{0,k}$, supplemented by the anomalous dimension $\eta = - \frac{1}{\bar Z_\phi} \partial_t \bar Z_\phi$. 

The flow equation for $\lambda_k$ is obtained by taking the derivative of $\partial_t U_k(\rho)$ in Eq.\ \eqref{eq:B6} according to
\begin{equation}
\partial_t \lambda_k = \frac{\partial^2}{\partial \rho^2} \partial_t U_k(\rho){\big |}_{\rho_0} = 2 \eta_\phi \lambda_k + \partial_t U_k^{\prime\prime}(\rho){\big |}_{\bar \rho_0}
\end{equation}
and the last term can be taken from \eqref{eq:B10}. In terms of the dimensionless coefficients $\tilde \gamma_1^2= \gamma_1^2/k^2$, $Z_1$, $\tilde \lambda = \lambda_k$, and $\tilde \rho_0 = \rho_{0,k}/k^2$ one has
\begin{equation}
\begin{split}
\partial_t \tilde \lambda = & 2 \eta_\phi \tilde \lambda + \frac{1}{2} {\bigg [} - 9 \tilde \lambda^2 \; I_2\left( 2\tilde \lambda \tilde \rho_0, \tilde \gamma_1^2, Z_1, 0,c,4 \right)\\
& - (N-1) \tilde \lambda^2\; I_2\left( 0,0,1,0,c,4 \right) {\bigg ]}.
\end{split}
\label{eq:floweqlambdadimless}
\end{equation}
The coefficient $c$ is a parameter of the infrared regulator function. A reasonable choice is $c\approx 1$ and the dependence of the final result on the precise value of $c$ can be used for a rough estimation of systematical errors connected with the truncation. 

Let us now come to the flow equation for $\rho_{0,k}$ or $\tilde \rho_0$. It is obtained from the condition
\begin{equation}
\frac{d}{dt} U^\prime(\rho_{0,k}) = \partial_t U_k^\prime(\rho_{0,k}){\big |}_{\rho_{0,k}} + U_k^{\prime\prime}(\rho_{0,k}) \partial_t \rho_{0,k} = 0,
\end{equation}
or
\begin{equation}
\begin{split}
\partial_t \rho_{0,k} = & - \frac{1}{\lambda_k} \partial_t U^\prime(\rho_{0,k}) {\big |}_{\rho_{0,k}}\\
= & - \eta_\phi\, \rho_{0,k} - \frac{1}{\lambda_k} \partial_t U^\prime(\rho_{0,k}){\big |}_{\bar \rho_{0,k}}.
\end{split}
\end{equation}
In the last equation we have used \eqref{eq:B6}. For the dimensionless combination $\tilde \rho_0 = \rho_{0,k} / k^2$ this results in
\begin{equation}
\begin{split}
\partial_t \tilde \rho_0 = & -(2+\eta_\phi) \tilde \rho_0 - \frac{1}{2} {\bigg [} 3 I_1\left( 2\tilde \lambda \tilde \rho_0, \tilde \gamma_1^2, Z_1, 0, c, 4 \right)\\
& +(N-1) I_1\left( 0,0,1,0, c,4\right) {\bigg ]}.
\end{split}
\label{eq:floweqrhodimless}
\end{equation}
To solve Eqs.\ \eqref{eq:floweqlambdadimless} and \eqref{eq:floweqrhodimless} one needs to specify some initial values for $\tilde \lambda$ and $\tilde \rho$ at some large value of the infrared cutoff parameter $k=\Lambda$ or, equivalently, at $t=\ln(k/\Lambda) = 0$. One also needs the anomalous dimension $\eta_\phi$ and the coefficients $Z_1$ and $\gamma_1^2$ as a function of $t$. Equations that govern these will be discussed in the subsequent section.

\subsection{Flow of the propagator}\label{ssec:Propagator}
Let us now discuss the ($k$-dependent) propagator and its flow equation within our formalism. For a constant background field $\phi_1=\phi_0 = \sqrt{2\rho_0}$, $\phi_2=\dots = \phi_N=0$, the propagator is a diagonal matrix in momentum space as well as in the space of ``internal'' or $O(N)$ degrees of freedom labeled by the index $M$. 

The $k$-dependent, renormalized propagator of the radial field $\phi_1$ is given in the truncation \eqref{eq:Gamma_2truncren} by
\begin{equation}
G_1=\frac{1}{Z_1(-p_0^2+\vec p^2) - i s(p_0) \gamma_1^2 + 2 \lambda_k \rho_{0,k}}
\end{equation}
while the propagator of the Goldstone modes $\phi_2, \dots, \phi_N$ is
\begin{equation}
G_2 = \frac{1}{-p_0^2+\vec p^2}.
\end{equation}
To trace the $k$-dependence of these objects one needs besides the flow equations for $\lambda_k$ and $\rho_{0,k}$ also the ones for the discontinuity $\gamma_1^2$ and the coefficient $Z_1$ as well as the anomalous dimension $\eta_\phi= - \frac{1}{\bar Z_\phi} \partial_t \bar Z_\phi$. To derive these flow equations we expand in the fields according to \eqref{eq:expansionaroundconstantfields}. The term quadratic in the ``fluctuating fields'' $\delta\bar \phi$ is in general of the form
\begin{equation}
\Gamma_{k,2} = \int_{p_0,\vec p} \left\{ \frac{1}{2} \delta\bar\phi_1(-p_0,-\vec p) \bar P_1(p_0,\vec p) \delta \bar\phi_1(p_0,\vec p) + \sum_{M=2}^N \frac{1}{2} \delta\bar\phi_M(-p_0,-\vec p) \bar P_2(p_0,\vec p) \delta\bar\phi_M(p_0,\vec p) \right\}.
\end{equation}
Due to the $O(N)$ symmetry, the functions $\bar P_1(p_0,\vec p)$ and $\bar P_2(p_0,\vec p)$ can differ only for $\rho>0$. Within the truncation \eqref{eq:truncationbar} their particular form is given by \eqref{eq:Gamma_2trunc}. We make now an approximation where momentum- and frequency dependent parts proportional to to $Z_\rho$ or $\gamma_\rho^2$  are neglected in the effective vertices for the fields $\phi_1$ and $\phi_2$ on the right hand side of the flow equation. Using standard methods it is then straight forward to derive the flow equations
\begin{equation}
\begin{split}
\frac{1}{\bar Z_\phi}\partial_t & \frac{1}{2 q_0} \frac{\partial}{\partial q_0}  \bar P_1(q_0,0) = \\
& - \left[ 18 \rho (U^{\prime\prime})^2 + 24 \rho^2 U^{\prime\prime} U^{(3)} + 8 \rho^3 (U^{(3)})^2 \right] k^{d-6} \\
& \times J\left( \frac{q_0}{k}, \frac{U^\prime + 2 \rho U^{\prime\prime}}{k^2},\frac{U^\prime + 2 \rho U^{\prime\prime}}{k^2}, \frac{\gamma_1^2}{k^2}, \frac{\gamma_1^2}{k^2},Z_1,Z_1, \frac{T}{k}, c, d \right)\\
& - 2 (N-1)^2 \rho (U^{\prime\prime})^2 k^{d-6} \; J\left( \frac{q_0}{k}, \frac{U^\prime }{k^2},\frac{U^\prime }{k^2}, 0, 0,1,1, \frac{T}{k}, c, d \right),\\
\frac{1}{\bar Z_\phi}\partial_t & \frac{1}{2 q_0} \frac{\partial}{\partial q_0} \bar P_2(q_0,0)  = \\
& - 4 (N-1) \;\rho (U^{\prime\prime})^2 k^{d-6} J\left( \frac{q_0}{k}, \frac{U^\prime + 2 \rho U^{\prime\prime}}{k^2},\frac{U^\prime}{k^2}, \frac{\gamma_1^2}{k^2}, 0,Z_1,1, \frac{T}{k}, c, d \right),\\
\frac{1}{\bar Z_\phi} \partial_t & \text{Disc}_{q_0} \bar P_1(q_0,0) \\
& = \left[ 18 \rho (U^{\prime\prime})^2 + 24 \rho^2 U^{\prime\prime} U^{(3)} + 8 \rho^3 (U^{(3)})^2 \right] k^{d-6} \\
& \times K\left( \frac{q_0}{k}, \frac{U^\prime + 2 \rho U^{\prime\prime}}{k^2},\frac{U^\prime + 2 \rho U^{\prime\prime}}{k^2}, \frac{\gamma_1^2}{k^2}, \frac{\gamma_1^2}{k^2},Z_1, Z_1, \frac{T}{k}, c, d \right)\\
&+2 (N-1)^2 \rho (U^{\prime\prime})^2 k^{d-6} \; K\left( \frac{q_0}{k}, \frac{U^\prime }{k^2},\frac{U^\prime }{k^2}, 0, 0,1,1, \frac{T}{k}, c, d \right),\\
\frac{1}{\bar Z_\phi} \partial_t & \text{Disc}_{q_0} \bar P_2(q_0,0)\\
& =  4 (N-1) \rho (U^{\prime\prime})^2 k^{d-6} K\left( \frac{q_0}{k}, \frac{U^\prime + 2 \rho U^{\prime\prime}}{k^2},\frac{U^\prime}{k^2}, \frac{\gamma_1^2}{k^2}, 0, Z_1, 1,\frac{T}{k}, c, d \right).
\end{split}
\label{eq:C2}
\end{equation}
We used here the integral functions $J$ and $K$ as defined in appendix \ref{app:integralfunctions} (also the operator $\text{Disc}$ is defined there). From these equations one can easily infer the anomalous dimension $\eta_\phi = -\tfrac{1}{\bar Z_\phi} \partial_t \bar Z_\phi$ according to
\begin{equation}
\begin{split}
\eta_\phi = & - \frac{1}{\bar Z_\phi} \partial_t \bar Z_\phi = \frac{1}{\bar Z_\phi} \partial_t \frac{1}{2 q_0} \frac{\partial}{\partial q_0} \bar P_2(q_0,0) {\big |}_{q_0=0}\\
 = & - 4 (N-1) \tilde \rho \tilde \lambda^2\; J\left( 0, 2 \tilde \lambda \tilde \rho_0,0,\tilde \gamma_1^2 ,0,Z_1,1,0,c,4 \right).
\label{eq:anomalousdim}
\end{split}
\end{equation}
In a similar way one can obtain the flow equation for $Z_1$ from \eqref{eq:C2}
\begin{equation}
\begin{split}
\partial_t Z_1 = & \eta_\phi Z_1 - \frac{1}{\bar Z_\phi} \frac{1}{2 q_0} \frac{\partial}{\partial q_0} \bar P_1(q_0,0) {\big |}_{q_0=m_1}\\
=& \eta_\phi Z_1 - 18 \; \tilde \rho_0 \tilde \lambda^2 \; J\left( \tilde q_0, 2\tilde \lambda \tilde \rho_0, 2\tilde \lambda \tilde \rho_0, \tilde \gamma_1^2, \tilde \gamma_1^2, Z_1, Z_1, 0,c,4\right)\\
& + 2 (N-1)^2 \; \tilde \rho_0 \tilde \lambda^2 J\left(\tilde q_0,0,0,0,0,1,1,0,c,4 \right).
\end{split}
\label{eq:5.12}
\end{equation}
According to the definition in \eqref{eq:defZ1gamma1} this should be evaluated as $\tilde q_0 =m_1/k =\sqrt{2\tilde\lambda \tilde \rho_0 / Z_1}$. 
The flow equation for the discontinuity $\tilde\gamma_1^2$ can be obtained from \eqref{eq:C2} as
\begin{equation}
\begin{split}
\partial_t \tilde \gamma_1^2 = & (\eta_\phi - 2) \tilde \gamma_1^2 + \frac{1}{\bar Z_\phi k^2} \partial_t \text{Disc}_{q_0} \bar P_2(q_0,0) {\big |}_{q_0=m_1}\\
= & (\eta_\phi-2) \tilde \gamma_1^2 \\
&+ 18 \; \tilde \lambda^2 \tilde \rho_0 \; K\left( \tilde q_0, 2 \tilde \lambda \tilde \rho_0, 2 \tilde \lambda \tilde \rho_0, \tilde \gamma_1^2, \tilde \gamma_1^2, Z_1, Z_1, 0, c, 4 \right)\\
& + 2 (N-1)^2 \tilde \lambda^2 \tilde \rho_0\; K \left( \tilde q_0, 0,0,0,0,1,1,0,c,4 \right).
\end{split}
\label{eq:5.14}
\end{equation} 
Again this should be evaluated at $\tilde q_0 = \sqrt{2\tilde \lambda \tilde \rho_0/Z_1}$. The second term on the right hand side of \eqref{eq:5.14}, which is due to fluctuations of the massive radial field, will not contribute to the flow of the discontinuity since the decay of a massive particle into two particles of the same mass is kinematically not possible. However, this is different for the last term in \eqref{eq:5.14} which is due to fluctuations of the Goldstone modes. A massive particle can decay into two massless ones so that a non-zero decay width of the radial field is generated by this term.

\subsection{Numerical results}\label{sec:Numericalresults}
\begin{figure}
\centering
\begin{picture}(240,170)
\put(0,8){\includegraphics[scale=0.98]{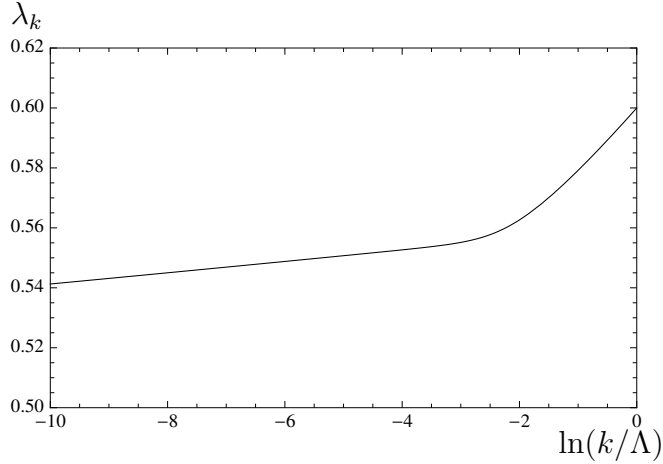}}
\put(0,162){$\lambda_k$}
\put(205,0){$\ln (k/\Lambda)$}
\end{picture}
\caption{Flow of the interaction strength $\lambda_k$.}
\label{fig:flowlambda}
\end{figure}

\begin{figure}
\centering
\begin{picture}(240,170)
\put(0,8){\includegraphics[scale=0.98]{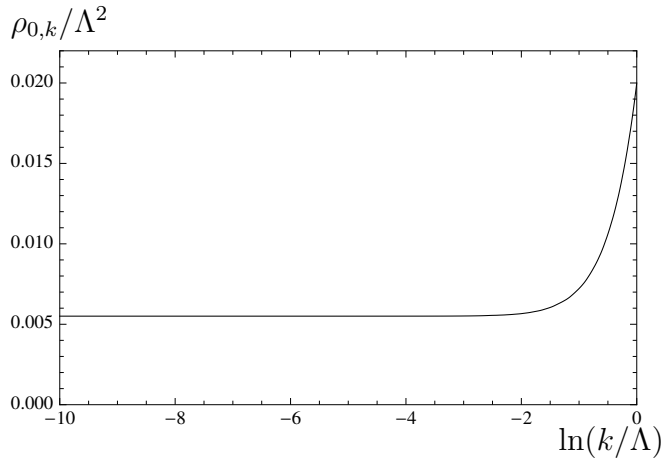}}
\put(0,158){$\rho_{0,k}/\Lambda^2$}
\put(205,0){$\ln (k/\Lambda)$}
\end{picture}
\caption{Flow of the minimum of the effective potential $\rho_{0,k}$.}
\label{fig:flowrho}
\end{figure}

\begin{figure}
\centering
\begin{picture}(240,170)
\put(0,8){\includegraphics[scale=0.98]{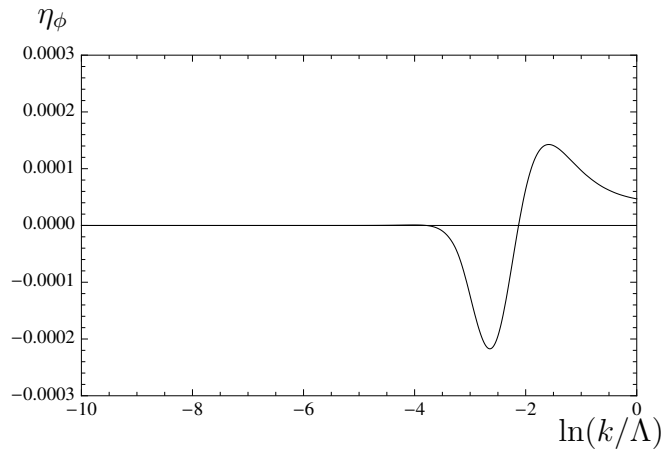}}
\put(10,158){$\eta_\phi$}
\put(205,0){$\ln (k/\Lambda)$}
\end{picture}
\caption{Anomalous dimension $\eta_\phi$ as obtained from \eqref{eq:anomalousdim}.}
\label{fig:ad}
\end{figure}

\begin{figure}
\centering
\begin{picture}(240,170)
\put(0,8){\includegraphics[scale=0.98]{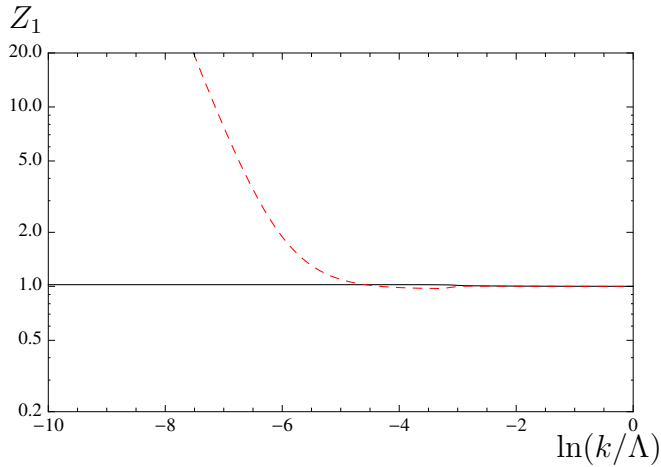}}
\put(0,162){$Z_1$}
\put(205,0){$\ln (k/\Lambda)$}
\end{picture}
\caption{Flow of the coefficient $Z_1$ as obtained from \eqref{eq:5.12} (solid line). We also show the resulting behavior if the flow equation is evaluated at $q_0=0$ instead (dashed line). Interestingly, one finds $Z_1\to \infty$ for $k\to 0$ in the latter case whereas the result is completely regular if the flow equation is evaluated on-shell.}
\label{fig:flowZ1}
\end{figure}

\begin{figure}
\centering
\begin{picture}(240,170)
\put(0,8){\includegraphics[scale=0.98]{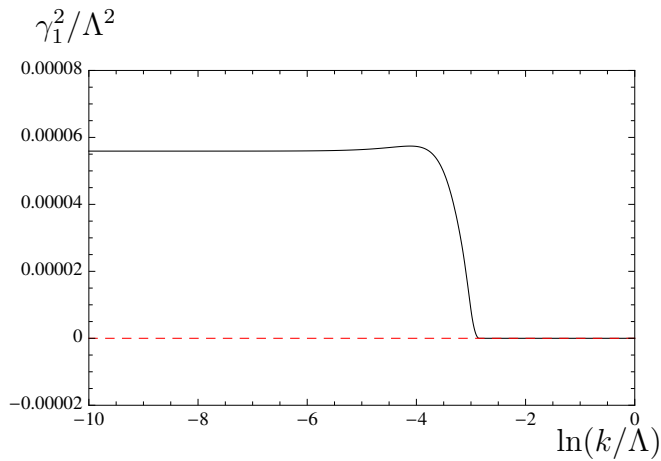}}
\put(10,157){$\gamma_1^2/\Lambda^2$}
\put(205,0){$\ln (k/\Lambda)$}
\end{picture}
\caption{Flow of the discontinuity coefficient $\gamma_1^2$ as obtained from \eqref{eq:5.14} (solid line). We also show the resulting behavior if the flow equation is evaluated at $q_0=0$ instead (dashed line). As expected, the discontinuity $\gamma_1^2$ is non-zero on-shell whereas it vanishes for $q_0=0$.}
\label{fig:flowgamma1}
\end{figure}

Equations \eqref{eq:floweqlambdadimless}, \eqref{eq:floweqrhodimless}, \eqref{eq:anomalousdim}, \eqref{eq:5.12} and \eqref{eq:5.14} constitute a closed set of flow equations which can be solved numerically. For doing that one also needs to specify some initial values at the UV scale $k=\Lambda$, of course. For an illustration we choose them here to be
\begin{equation}
\tilde \lambda(\Lambda) = 0.6, \quad \tilde \rho_0(\Lambda)=0.02, \quad Z_1(\Lambda) = 1\quad \text{and}\quad \tilde \gamma_1^2(\Lambda)=0.
\end{equation}
The resulting flow behavior for the ${\cal O}(N)$ model in $3+1$ space-time dimensions is shown in Figs.\ \ref{fig:flowlambda} to \ref{fig:flowgamma1}.

In Fig.\ \ref{fig:flowlambda} we show the (rather weak) scale dependence of the interaction strength $\lambda$.
As expected, the dependence on $k$ is logarithmic. Strictly speaking, the logarithmic running implies $\lambda(k)\to 0$ for $k\to 0$ which is closely connected to the ``triviality'' of the theory. One can clearly separate two regimes in Fig.\ \ref{fig:flowlambda}. For large scales $k$ the logarithmic running is stronger since both fluctuations of the Goldstone modes and of the radial mode contribute. At smaller scales $k$ the contribution of the radial mode becomes small since it is suppressed by its non-zero mass $m_1=\sqrt{2\lambda_k \rho_{0,k}/Z_1}$. The transition between the two regimes takes place for $m_1\approx k$. 

In Fig.\ \ref{fig:flowrho} we show the $k$-dependence of the minimum of the effective potential $\rho_{0,k}$. For large scale parameters it first decreases when $k$ is lowered before it settles to a finite value where it remains for $k\to 0$. 
In Fig.\ \ref{fig:ad} we show the anomalous dimension $\eta_\phi = - \frac{1}{\bar Z_\phi}\partial_t \bar Z_\phi$ as a function of the scale. For large $k$ it has an oscillating behavior before it goes to zero for smaller $k$.

More interesting for the purpose of the present paper is the flow of the coefficient $Z_1$ and the discontinuity or decay coefficient $\gamma_1^2$ as shown in Fig.\ \ref{fig:flowZ1} and \ref{fig:flowgamma1}, respectively. We show the solution of the flow equation evaluated at the frequency $q_0=m_1$ as discussed in the preceding section (solid lines) but also the corresponding result if the flow equations for $Z_1$ and $\gamma_1^2$ are evaluated at vanishing frequency $q_0=0$ instead (dashed lines). 
Interestingly, $Z_1$ changes only very little from $Z_1=1$ to a slightly larger value if the flow equation is evaluated at $q_0=m_1$ but diverges for $k\to 0$ if evaluated at $q_0=0$.

As expected, the flow of $\gamma_1^2$ vanishes identically when evaluated at $q_0=0$ but is non-trivial at $q_0=m_1$. One finds that $\gamma_1^2$ starts to deviate from zero at some scale $k \approx q_0$, increases then strongly before it decreases again slightly and settles to some finite positive value where it remains for $k\to 0$. The physical origin of this discontinuity in the propagator of the radial mode is that it can decay into two massless Goldstone excitations. We emphasize that the present calculation takes this decay width in a self-consistent manner into account.

\section{Conclusions}\label{sec:conclusions}
We have discussed how functional renormalization group equations can be analytically continued from imaginary Matsubara frequencies to the real frequency axis. For the specific example of a relativistic scalar field with ${\cal O}(N)$ symmetry we have derived flow equations for the propagator of the radial mode that are evaluated for real frequencies corresponding to the on-shell energy of this massive excitation. A prominent and interesting feature is the imaginary discontinuity of the inverse propagator which is closely connected to the particle decay width.

An improved derivative expansion formalism allows to expand in frequencies and momenta around their on-shell values in Minkowski space. The thus approximated propagators can be used in a self-consistent way in loop expressions on the right hand side of flow equations if standard methods based on complex contour integration are used to perform the Matsubara frequency summations. This adapted version of the derivative expansion in Minkowski space is very close to the physical dynamics. Since the dynamical properties of an excitation are dominated to a large extend by the singular part of its propagator one can expect that the convergence properties are better compared to a derivative expansion in Euclidean space.

We emphasize that our formalism fully conserves Lorentz symmetry. This is facilitated by a particular choice of infrared regulator function which decays algebraically both for large spatial momenta and large imaginary frequencies. Its rather simple form allows to perform all frequency summations or integrations analytically but still leads to expressions for flow equations that are ultraviolet and infrared finite and thus need no further regularization.

Although we have discussed here for simplicity of notation only relativistic scalar fields, the formalism can be extended in a straight-forward way to more complicated theories with bosonic and fermionic fields of different spin at arbitrary temperature and chemical potential as well as to non-relativistic field theories.

The formation of bound states or other composite degrees of freedom constitutes a particular interesting field of application. Using a recently derived exact flow equation for composite operator fields it is possible to change between a description in terms of fundamental and composite fields in a continuous manner during the renormalization group evolution \cite{Floerchinger:2009uf}, see also \cite{Gies:2001nw, Floerchinger:2010da}.
So far, this program was limited to some extend by the difficulty to identify for a given correlation function the part that is due to the exchange of some composite degree of freedom. The formalism developed in the present paper will allow to determine this part directly from the on-shell singularity for the corresponding real frequency. This is in general not possible without the analytic continuation from Matsubara space.

Another interesting prospect for the formalism is the calculation of transport properties. Indeed, all quantities that are accessible from linear response theory such as viscosities, conductivities, permittivities, permeabilities etc.\ or the corresponding relaxation times can now be calculated from functional renormalization. The flow equations of the static and dynamic response functions can be determined within a given truncation. In some cases it may be advantageous to introduce appropriate composite degrees of freedom explicitly into the formalism.

In summary, we believe that the analytic continuation of functional renormalization group equations brings the formalism closer to the physical dynamics in Minkowski space, allows to calculate more observables and could finally lead to more accurate results with comparatively little computational effort.

\begin{appendix}
\section{Integral functions}\label{app:integralfunctions}
We define the integral functions according to
\begin{equation}
\begin{split}
& I_{n}(\tilde m^2,\tilde \gamma^2, z, \tilde T, c, d) =\\
& k^{2n-d}(\delta_{0n}-n) \int_{p_0} \int \frac{d^{d-1} p}{(2\pi)^{d-1}}  \frac{1}{\left[ z(-p_0^2+\vec p^2)+\tilde m^2 k^2-i s(p_0) \tilde \gamma^2 k^2+\tfrac{1}{Z} R_k \right]^{n+1}}\tfrac{1}{Z} \partial_t R_k
\end{split}
\label{eq:intfunctI}
\end{equation}
where the frequency $p_0$ is summed over the discrete imaginary values $p_0 = 2 \pi i T n $ with $\int_{p_0} = T \sum_n$, $T=\tilde T k$ and $t=\ln (k/\Lambda)$. Since the cutoff function $R_k$ is usually proportional to the wavefunction renormalization constant $Z$, the right hand side of \eqref{eq:intfunctI} includes a term proportional to $\eta=-\tfrac{1}{Z}\partial_t Z$. In many situations one has $\eta\ll 1$ and this correction can therefore often be neglected.

Note that in terms of a formal cutoff derivative $\tilde \partial_t$ which hits only the cutoff $R_k$, one can write for $n\geq 1$
\begin{equation}
I_{n}(\tilde m^2,\tilde \gamma^2, z,\tilde T, c, d) = k^{2n-d}\tilde \partial_t \int_{p_0} \int \frac{d^{d-1} p}{(2\pi)^{d-1}}  \frac{1}{\left[ z(-p_0^2+\vec p^2)+\tilde m^2 k^2-i s(p_0) \tilde \gamma^2 k^2+\tfrac{1}{Z} R_k \right]^{n}}.
\end{equation}
We note also the recursion relation 
\begin{equation}
\frac{\partial}{\partial \tilde m^2} I_n(\tilde m^2,\tilde \gamma^2, z, \tilde T, c, d) = (\delta_{0n}-n)\; I_{n+1}(\tilde m^2,\tilde \gamma^2, z, \tilde T, c, d).
\end{equation}

For the flow equations of the propagator one needs also the integral functions
\begin{equation}
\begin{split}
J(\tilde q_0,\tilde m_1^2,\tilde m_2^2, \tilde \gamma_1^2,\tilde \gamma_2^2, z_1, z_2, \tilde T, c,d) = & - k^{4-d} \tilde \partial_t \;\frac{1}{2\tilde q_0} \frac{\partial}{\partial \tilde q_0} \int_{p_0} \int \frac{d^{d-1}p}{(2\pi)^{d-1}}\\
&\times \frac{1}{[z_1(-p_0^2+\vec p^2)+ \tilde m_1^2 k^2-i s(p_0) \tilde \gamma_1^2 k^2+\tfrac{1}{Z}R_k]}\\
& \times \frac{1}{[z_2\left(-(p_0+\tilde q_0 k)^2+\vec p^2\right)+\tilde m_2^2 k^2-i s(p_0+q_0) \tilde \gamma_2^2 k^2 + \tfrac{1}{Z}R_k]}
\end{split}
\label{eq:intfuncJ}
\end{equation}
and
\begin{equation}
\begin{split}
K(\tilde q_0,\tilde m_1^2,\tilde m_2^2, \tilde \gamma_1^2, \tilde \gamma_2^2, z_1, z_2, c, d) = & k^{4-d} \tilde \partial_t\; \text{Disc}_{\tilde q_0} \int_{p_0} \int \frac{d^{d-1}p}{(2\pi)^{d-1}}\\
&\times \frac{1}{[z_1(-p_0^2+\vec p^2)+ \tilde m_1^2 k^2-i s(p_0) \tilde \gamma_1^2 k^2+\tfrac{1}{Z}R_k]}\\
& \times \frac{1}{[z_2\left(-(p_0+\tilde q_0 k)^2+\vec p^2\right)+\tilde m_2^2 k^2-i s(p_0+q_0) \tilde \gamma_2^2 k^2 + \tfrac{1}{Z}R_k]}
\label{eq:defintfunctK}
\end{split}
\end{equation}
where the operator $\text{Disc}$ is defined (for $\tilde q_0>0$) by
\begin{equation}
\text{Disc}_{\tilde q_0} f(\tilde q_0) = \frac{i}{2} \left[ f(\tilde q_0+i \epsilon) - f(\tilde q_0-i \epsilon) \right].
\end{equation}
For $\tilde q_0<0$ the definition of the operator $\text{Disc}$ contains an additional minus sign.
In praxis the integral in \eqref{eq:defintfunctK} has for $k>0$ not only a single discontinuity on the real frequency axis but several ones on lines that are approximately parallel to the real $q_0$ axis. They merge only for $k\to 0$. Instead of tracing the complete analytic structure, we follow in this paper an approximation scheme where the $k$-dependent inverse propagator is taken to have only a single cut for real $q_0$. The flow equation of this discontinuity is determined by summing the different contributions in Eq.\ \eqref{eq:defintfunctK}. This procedure gives the correct result for $k\to 0$ and for intermediate $k$ it should be a reasonable approximation.

Using the regulator function given in section \ref{ssec:Regulator} and the expression for the regularized propagator given there, one can use standard methods to perform the Matsubara frequency summation in Eqs.\ \eqref{eq:intfunctI}, \eqref{eq:intfuncJ} and \eqref{eq:defintfunctK}. The integrals along branch cuts get substantially simplified when one assumes that they are dominated by the poles that are close to them on the different Riemann sheets and can then be done analytically. Some of the remaining integrals over spatial momenta can be done analytically as well, others numerically.

\end{appendix}

\medskip
\section*{Acknowledgments}
\medskip
Financial support by DFG under contract FL 736/1-1 is gratefully acknowledged.


\begin{thebibliography}{}

\bibitem{Wilson}
K.~G.~Wilson, Phys.\ Rev.\ B {\bf 4}, 3174 (1971).

\bibitem{WegnerHoughton}
F.~J.~Wegner and A.~Houghton, Phys.\ Rev.\ A {\bf 8}, 401 (1973).

\bibitem{Polchinski}
J.~Polchinski, Nucl.\ Phys.\ B {\bf 231}, 269 (1984).

\bibitem{Wetterich1993}
C.~Wetterich, Phys.\ Lett.\ B {\bf 301}, 90 (1993).


\bibitem{FRGReviews}
J.~Berges, N.~Tetradis and C.~Wetterich, Phys.\ Rep.\ {\bf 363}, 223 (2002);
T.~R.~Morris, Prog.\ Theor.\ Phys.\ Suppl.\  {\bf 131}, 395 (1998);
K.~Aoki, Int.\ J.\ Mod.\ Phys.\ B {\bf 14}, 1249 (2000); 
C.~Bagnuls and C.~Bervillier, Phys.\ Rept.\  {\bf 348}, 91 (2001); 
M.~Salmhofer and C.~Honerkamp, Prog.\ Theor.\ Phys.\  {\bf 105}, 1 (2001); 
J.~Polonyi, Central Eur.\ J.\ Phys.\  {\bf 1}, 1 (2003); 
W.~Metzner, Prog.\ Theor.\ Phys.\ Suppl.\ {\bf 160}, 58 (2005);
H.~Gies, hep-ph/0611146;
B.~Delamotte, cond-mat/0702365;
M.~Niedermaier and M.~Reuter, Living Reviews in Relativity {\bf 9}, 5 (2006);
J.~M.~Pawlowski, Annals Phys. {\bf 322}, 2831 (2007);
B.~J.~Schaefer and J.~Wambach, Phys. Part. Nucl.{\bf 39}, 1025 (2008);
O.~J.~Rosten, arXiv:1003.1366;
M.~M.~Scherer, S.~Floerchinger and H.~Gies, Phil. Trans. R. Soc. A {\bf 369}, 2779 (2011);
J.~Braun, arXiv:1108.4449.


\bibitem{Canet}
L.~Canet, B.~Delamotte, O.~Deloubriere and N.~Wschebor, Phys.\ Rev.\ Lett.\ {\bf 92}, 195703 (2004);
L.~Canet, H.~Chant\'e and B.~Delamotte, Phys.\ Rev.\ Lett.\ {\bf 92}, 255703 (2004);
L.~Canet, H.~Chant\'e and B.~Delamotte, I.~Dornic and M.~A.~Mu\~noz, Phys.\ Rev.\ Lett.\ {\bf 95}, 100601 (2005).

\bibitem{Gasenzer:2008zz} 
  T.~Gasenzer and J.~M.~Pawlowski,
  Phys.\ Lett.\ B {\bf 670}, 135 (2008);
  T.~Gasenzer, S.~Kessler and J.~M.~Pawlowski,
  Eur.\ Phys.\ J.\ C {\bf 70}, 423 (2010).

\bibitem{Berges:2008sr} 
  J.~Berges and G.~Hoffmeister,
  Nucl.\ Phys.\ B {\bf 813}, 383 (2009).

\bibitem{Manrique:2011jc} 
  E.~Manrique, S.~Rechenberger and F.~Saueressig,
  Phys.\ Rev.\ Lett.\  {\bf 106}, 251302 (2011).

\bibitem{Zerodimensional}
S.~Kehrein, Phys.\ Rev.\ Lett.\ {\bf 95}, 056602 (2005);
G.~Gezzi, T.~Pruschke and V.~Meden, Phys.\ Rev.\ B {\bf 75}, 045324 (2007);
S.~G.~Jakobs, V.~Meden and H.~Schoeller, Phys.\ Rev.\ Lett.\ {\bf 99}, 150603 (2007);
T.~Korb, F.~Reininghaus, H.~Schoeller and J.~K\"onig, Phys.\ Rev.\ B {\bf 76}, 165316 (2007);
C.~Karrasch, R.~Hedden, R.~Peters, T.~Pruschke, K.~Schšnhammer, and V.~Meden, 
J. Phys.: Condensed Matter {\bf 20}, 345205 (2008);
H.~Schoeller and F. ~Reininghaus, Phys.\ Rev.\ B {\bf 80}, 045117 (2009); 
H.~Schoeller, Eur. Phys. J. Special Topics {\bf 168}, 179 (2009);
M.~Pletyukhov, D.~Schuricht and H.~Schoeller, Phys.\ Rev.\ Lett.\ {\bf 104}, 106801 
(2010);
S.~G.~Jakobs, M.~Pletyukhov and H.~Schoeller, Phys.\ Rev.\ B {\bf 81}, 195109 (2010);
C.~Karrasch, M.~Pletyukhov, L.~Borda and V.~Meden, Phys.\ Rev.\ B {\bf 81}, 125122 (2010);
S.~Andergassen, M.~Pletyukhov, D.~Schuricht, H.~Schoeller and L.~Borda, Phys. Rev. B {\bf 83}, 205103 (2011);
D.~M.~Kennes, S.~G.~Jakobs, C.~Karrasch and V.~Meden, e-print arXiv:1111.6982.

\bibitem{Dupuis2009}
N.~Dupuis, Phys.\ Rev.\ A {\bf 80}, 043627 (2009).

\bibitem{Sinner:2009zz} 
  A.~Sinner, N.~Hasselmann and P.~Kopietz,
  Phys.\ Rev.\ Lett.\  {\bf 102}, 120601 (2009)

\bibitem{SchmidtEnss}
R.\ Schmidt, T.\ Enss, Phys.\ Rev.\ A {\bf 83}, 063620 (2011).

\bibitem{BMW}
J.~P.~Blaizot, R.~Mendez-Galain and N.~Wschebor, Phys.\ Rev.\ E {\bf 74}, 051116 (2006); Phys.\ Rev.\ E {\bf 74}, 051117(2006); Phys.\ Lett.\ B {\bf 632}, 571 (2006); Eur.\ Phys.\ J.\ B {\bf 58}, 297 (2007).

\bibitem{BMWScalar}
  J.~P.~Blaizot, A.~Ipp, R.~Mendez-Galain and N.~Wschebor,
  Nucl.\ Phys.\ A {\bf 784}, 376 (2007);
  J.~P.~Blaizot, A.~Ipp and N.~Wschebor,
  Nucl.\ Phys.\ A {\bf 849}, 165 (2011).

\bibitem{Coleman85}
S. Coleman, {\itshape Aspects of Symmetry}, (Cambridge University Press, Cambridge, 1985).

\bibitem{Weinberg96}
S. Weinberg, {\itshape The Quantum theory of fields}, (Cambridge University Press, Cambridge, 1996).

\bibitem{Litim2001}
D. F. Litim, Phys. Rev. D {\bf 64}, 105007 (2001). 

\bibitem{Floerchinger:2009uf} 
  S.~Floerchinger and C.~Wetterich,
  Phys.\ Lett.\ B {\bf 680}, 371 (2009).
  
\bibitem{Gies:2001nw} 
  H.~Gies and C.~Wetterich,
  Phys.\ Rev.\ D {\bf 65}, 065001 (2002);
  Acta Phys.\ Slov.\  {\bf 52}, 215 (2002).

\bibitem{Floerchinger:2010da} 
  S.~Floerchinger,
  Eur.\ Phys.\ J.\ C {\bf 69}, 119 (2010).
\end{thebibliography}
\end{document}